\documentclass[12pt,preprint]{aastex}
\usepackage{epsf}
\usepackage{lscape}

\usepackage{graphicx}
\usepackage{natbib}
\begin{document}

\title {TESTING WEAK LENSING MAPS WITH REDSHIFT SURVEYS: A SUBARU FIELD}

\author {Michael J. Kurtz} 
\affil{Smithsonian Astrophysical Observatory,
\\ 60 Garden St., Cambridge, MA 02138}
\email{mkurtz@cfa.harvard.edu}
\author {Margaret J. Geller} 
\affil{Smithsonian Astrophysical Observatory,
\\ 60 Garden St., Cambridge, MA 02138}
\email{mgeller@cfa.harvard.edu}
\author{Yousuke Utsumi}
\affil{The Graduate University for Advanced Studies, 
\\ 2-21-1 Osawa, Mitaka, Tokyo, Japan}
\affil {National Astronomical Observatory of Japan}
\email {yousuke.utsumi@nao.ac.jp}
\author{Satoshi Miyazaki}
\affil {National Astronomical Observatory of Japan,
\\ 2-21-1 Osawa, Mitaka, Tokyo, Japan}
\email{satoshi@subaru.naoj.org}
\author {Ian P. Dell'Antonio} 
\affil{Department of Physics, Brown University, 
\\Box 1843, Providence, RI 02912}
\email{ian@het.brown.edu}
\author {Daniel G. Fabricant} 
\affil{Smithsonian Astrophysical Observatory,
\\ 60 Garden St., Cambridge, MA 02138}
\email{dfabricant@cfa.harvard.edu}

\begin{abstract}

We use a dense redshift survey in the foreground of the Subaru GTO2deg$^2$ weak lensing field (centered at $\alpha_{2000}$ = 16$^h$04$^m$44$^s$;$\delta_{2000}$ =43$^\circ$11$^{\prime}$24$^{\prime\prime}$) to assess the completeness and comment on the purity of massive halo identification in the weak lensing map. The redshift survey (published here) includes 4541 galaxies; 4405 are new redshifts measured with the Hectospec on the MMT. Among the weak lensing peaks with a signal-to-noise greater that 4.25, 2/3 correspond to individual massive systems; this result is essentially identical to the Geller et al. (2010) test of the Deep Lens Survey field F2. The Subaru map, based on images in substantially better seeing than the DLS, enables detection of less massive halos at fixed redshift as expected. We demonstrate that the procedure adopted by Miyazaki et al. (2007) for removing some contaminated peaks from the weak lensing map  improves agreement between the lensing map and the redshift survey in the identification of candidate massive systems.

\end{abstract}


\maketitle

\section{Introduction}
\label{intro}

Weak lensing maps and redshift surveys are fundamental, complementary tools
of modern cosmology. A weak lensing map provides a  weighted ``picture''
of projected surface mass density; a redshift survey provides a
three-dimensional map of the galaxy distribution resolving structures along
the line of sight within the broad window ``imaged'' by the lensing map. 

Although galaxies are biased tracers of the mass distribution, comparison of a
weak lensing map with a foreground redshift survey covering the appropriate
range promises progress in resolving some of the issues limiting 
applications of weak lensing to the identification of mass-selected cluster samples. Here we focus on the use of
weak lensing for identifying clusters of galaxies. 

This paper is the second in a series comparing massive clusters identified in a dense foreground
redshift survey with significant weak lensing map convergence peaks. In our first paper
(Geller et al. 2010), we compared clusters identified in a Hectospec redshift survey
(SHELS: Geller et al. 2005)  with a Deep Lens Survey (DLS; Wittman et al. 2001) weak lensing map derived from deep
Kitt Peak 4-meter observations.  Here we describe a Hectospec redshift survey of the GTO2deg$^2$ field where Miyazaki  et al. (2002, 2007) used  Subaru Telescope observations as the basis for their weak lensing convergence map. These two assessments of the efficacy of weak lensing for massive cluster identification provide insights into both the strengths and limitations of the technique.

Since the first identifications of rich clusters from weak lensing maps (Wittman et al. 2001; Miyazaki et al. 2002), the number of 
detections has risen steeply (Hetterscheidt
et al. 2005; Wittman et al. 2006; Schirmer et al. 2007;
Gavazzi \& Soucail 2007; Miyazaki et al. 2007; Maturi et al. 2007; Berg\'e et al. 2008; Dietrich et al. 2008; Bellagamba et al. 2011; Shan et al. 2011). Geller et al. (2010) discuss the varying approaches and success rates of weak lensing surveys as rich cluster identifiers. Their comparison between clusters identified in the SHELS redshift survey and the DLS weak lensing map agrees with the more pessimistic observational and theoretical assessments of the technique (e.g. Hamana et al 2004; Hennawi \& Spergel 2005; Schirmer et al. 2007; Shan et al. 2011).  Geller et al. (2010) conclude that only 4 of the 12 significant DLS weak lensing peaks with signal-to-noise greater than 3.5 coincide with clusters massive enough to account for the lensing signal.

The SHELS/DLS weak-lensing purity for cluster detection is more pessimistic than many
previous analyses at least in part because Geller et al. (2010) require that the velocity dispersion
determined from the SHELS redshift survey be large enough to account for the lensing signal. 
Previous studies based on counts or photometric redshifts could not provide a velocity dispersion
and often they could not  distinguish between a genuine massive cluster and several groups embedded in the same large-scale structure or superimposed along the line-of-sight.

The underlying causes of the low purity of the DLS cluster candidate list are hard to identify from the single survey field Geller et al. (2010) studied. 
To gain further understanding of the relationship between clusters identified in a redshift survey
and significant peaks in a weak lensing convergence map, we carried out a redshift survey of a second field. Our goal is an independent measurement of the completeness and purity of weak lensing maps in
identifying massive clusters. 

To explore the comparison of foreground redshift surveys with weak lensing maps further, we chose a weak lensing map and candidate halo list constructed by different observers on a different facility. 
The Subaru weak lensing survey (Miyazaki et al. 2002; 2007) provides an interesting comparison with the DLS because
(1) Subaru is an 8.2-meter telescope , (2) the median seeing for the Subaru data was $\sim$0.66$^{\prime\prime}$ in contrast with the 0.85$^{\prime\prime}$ seeing for the DLS and (3) the algorithms 
for constructing the weak lensing map and for identifying significant peaks differ in the two projects. The typical number of sources in the Subaru map is 35 arcmin$^{-2}$; in the DLS map it is
22.4 arcmin$^{-2}$. At a fixed redshift within the lensing kernel of both surveys,  the deeper Subaru map with observations in better seeing should correspond more closely to the set of massive systems in the redshift survey. Detections should be at higher signal-to-noise for systems of fixed mass at a fixed redshift. The Subaru map should also enable detection of less massive systems at a fixed redshift. We test these expectations here. 

In Section \ref{data} we discuss the history of related observations
(Section \ref{history}) and we describe the redshift survey (Section \ref {redshifts}).
Our table of redshifts in the field includes 4405 new measurements with Hectospec
(Fabricant et al. 1998, 2005).
Next we describe the revised Subaru weak lensing map (Sections \ref{newimage} and \ref{newmap}) and its sensitivity (Section \ref{sensitivity}). We sample the redshift survey in Section \ref{probes} and identify
candidate clusters in the weak lensing map and in the redshift survey
(Section \ref{clusters}). We discuss the completeness and purity of the weak lensing map in identifying
systems of galaxies and compare it with our earlier results for the DLS field F2 in
Section \ref{comparison}. We conclude in Section \ref{conclusion}.

\section {The Data}
\label{data}

Miyazaki et al. (2002; 2007) used the 8.2 meter Subaru Telescope to carry out a weak lensing survey
of thirteen well-separated fields covering a total of 21.82 deg$^2$ . They used lensing convergence maps to identify 100 candidate halos with signal-to-noise $\nu > $ 3.7; ten of these candidates are
within the GTO2deg$^2$ field we examine here. This published sample motivates our choice for the limiting significance of the weak lensing peaks to be explored with the redshift survey.

In the Miyazaki et al. (2007)
GTO2deg$^2$ field (centered at $\alpha_{2000}$ = 16$^h$04$^m$44$^s$;$\delta_{2000}$ =43$^\circ$11$^{\prime}$24$^{\prime\prime}$), our foreground redshift survey includes
redshifts for 4549 galaxies and covers the entire 2.1 deg$^2$ Subaru field. We follow the approach of Geller et al. (2010) and use the redshift survey as a route to examining the completeness and purity of the weak lensing map in
identifying massive halos.

\subsection {The Subaru GTO2deg$^2$ Field: Imaging and Previous Spectroscopy}
\label{history}

During the commissioning phase of Suprime-Cam on Subaru, Miyazaki et al. (2002) imaged part of  a field studied years earlier by Gunn et al. (GHO: 1986). Here we review previous 
spectroscopic observations of clusters in this field.

In 1998, Postman et al. (1998) used the Keck to make both photometric and spectroscopic observations
of the GHO cluster Cl 1604 + 4304 which lies within the GTO2deg$^2$ field. The cluster is a centrally concentrated, rich system at a redshift $z = 0.8967$  with a rest-frame line-of-sight velocity dispersion of 1226 km s$^{-1}$ (based on 22 members). The mass is $\sim 4\times 10^{15}$ M$_\odot$. Lubin et al. (1998) carried out a morphological study of the cluster based on HST observations and later Lubin et al. (2004) used XMM-Newton observations to study its x-ray properties. Lubin et al. (2000) studied the clusters Cl1604 + 4304 and Cl 1604 + 4321 suggesting that they form a supercluster. The cluster Cl 1604 +4321 is at a redshift  $z = 0.924$ with a rest-frame line-of-sight velocity dispersion of 935 km s$^{-1}$ and a mass of $\sim 2 \times 10^{15}$ M$_\odot$. Because of their substantial redshift, these clusters are beyond the range of sensitivity of the Subaru weak lensing map (see Section \ref{newmap}).

In 1999, Dressler et al. (1999) included the cluster Cl1601+42 in their 
spectroscopic catalog of 10  distant rich clusters of galaxies. Girardi \& Mezzetti (2001) analyzed the redshift distribution for the 46 cluster members. They 
obtained a mean redshift of 0.54 and a rest frame line-of-sight velocity dispersion of 646$^{+84}_{-87}$ km s$^{-1}$. They compute a  
cluster mass of $\sim 2.5 \times 10^{14}$ M$_\odot$. This cluster is
within the sensitivity range of the weak lensing map and we compare these results with our own estimate in Section \ref{clusters}.

In contrast with the observations of  small patches of the
original GHO field geared toward studying individual clusters at 
reasonably high redshift, the wider field Miyazaki et al (2002) observation provides the basis for a weak lensing convergence map of the region. The weak lensing convergence map is sensitive to systems at lower redshifts ranging from $\sim 0.2$ to $\sim 0.7$. In Section \ref{clusters}, we show that the cluster Cl1601+42 (previously studied by Dressler (1999) and Girardi \& Mezzetti (2001)) appears as one of the two most significant peaks in the weak lensing map.

Hamana et al (2009) carried out multi-object spectroscopy along four lines-of-sight
in the GTO2deg$^2$ field.  They  observed with a single 6$^{\prime}$ diameter slit mask
in each of four patches and measured a total of
92 redshifts. On the basis of 15 of these galaxies coincident in redshift, SL J1602.8+4335 is a cluster at a redshift of 0.42 with a rest-frame line-of-sight velocity dispersion of 675 km s$^{-1}$; this cluster is the other of the two most significant
weak lensing peaks in the Miyazaki et al. (2007) map. We  compare this measurement with our redshift survey results in Section \ref{clusters} and Table \ref{tbl:VDisp}.

\subsection {The Hectospec Redshift Survey of the GTO2deg$^2$ Field}
\label{redshifts}
We investigate the association between convergence peaks and clusters (halos) in the galaxy distribution based on a densely sampled redshift survey.  Our goal is identification of the systems of galaxies 
that should produce a weak lensing signal. We thus 
seek to identify candidate systems in the redshift survey with rest frame 
line-of-sight velocity dispersion $\gtrsim 500$ km s$^{-1}$ consistent with the sensitivity of the Subaru map. 

We constructed the galaxy catalog from the Sloan Digital Sky Survey (SDSS, Adelman-McCarthy 2008) r-band galaxy list for the
GTO2deg$^2$ field. The field covers the region $240.05^\circ \leq \alpha_{2000} \leq
242.32^\circ$ and $42.55^\circ \leq \delta_{2000} \leq 43.83^\circ$. 

Because we seek to identify the massive systems of galaxies corresponding to the weak lensing peaks, we focus the redshift survey on the red galaxy population. These red objects preferentially populate clusters.
This procedure also suppresses the blue foreground populations; the weak lensing map is insensitive to systems at redshift less than $ z \simeq 0.05$
(see Figure \ref{fig:sensitivity.ps}).

We acquired spectra for the objects with the Hectospec 
(Fabricant et al. 1998, 2005), a 300-fiber robotic instrument mounted on the MMT from
February 1, 2009 to April 27, 2009. The Hectospec observation
planning software (Roll et al. 1998) 
enables  efficient acquisition of a pre-selected sample of galaxies. The software allows assignment of priorities as a function of galaxy properties including apparent magnitude, color, and central surface brightness. We discuss the priority by color below.

The spectra cover the wavelength range 3500 --- 10,000 \AA
\ with a resolution of $\sim$6 \AA. Exposure times ranged from 0.75 --- 1.5 hours.
We reduced the data with the standard Hectospec pipeline
(Mink et al. 2007) and derived redshifts with RVSAO (Kurtz \& Mink 1998) with
templates constructed for this purpose (Fabricant et al. 2005). 
Repeat  observations yield robust estimates of the median error  in $cz$ where $z$ is the redshift and $c$ is the speed of light. For emission line objects, the median error (normalized by $(1 + z)$) is 27 km s$^{-1}$; the median for absorption line objects
(again normalized by $(1 + z$)) is 37 km s$^{-1}$.
(Geller et al. 2010; Fabricant et al.  2005).

There are 4541 galaxies in the region with a measured redshift; our Hectospec observations provide  4405 of these and the SDSS provides 136. In addition, we measured Hectospec redshifts for 54 of the 136 SDSS objects. The SDSS objects are brighter and at lower redshift than the others we observed with Hectospec. Table 1 lists all of the
objects and their redshifts. The Table includes the SDSS ID (column 1), the J2000 right ascension (column 2), the J2000 declination (column 3), the SDSS Petrosian $r_{petro}$ magnitude (column 4), the SDSS fiber color $g-r$ (column 5), the SDSS fiber color $r-i$ (column 6), the
redshift (column 7), the redshift error (column 8), and the redshift source (column 9).

Our redshifts agree very well with  the 54 objects in common with SDSS; these
objects range in redshift from 0.019 to 0.468. At  lower redshifts, the galaxies in common are emission line objects and at higher redshift they are from the SDSS red galaxy sample. The overlapping objects are brighter than the typical galaxy in our sample. The mean difference $\delta{z}/(1 + z)$ = 3.4 km s$^{-1}$; the median difference is -0.8 km s$^{-1}$. The dispersion around the mean is 36.8 km s$^{-1}$. 

\begin{figure}
\centerline{\includegraphics[width=7.0in]{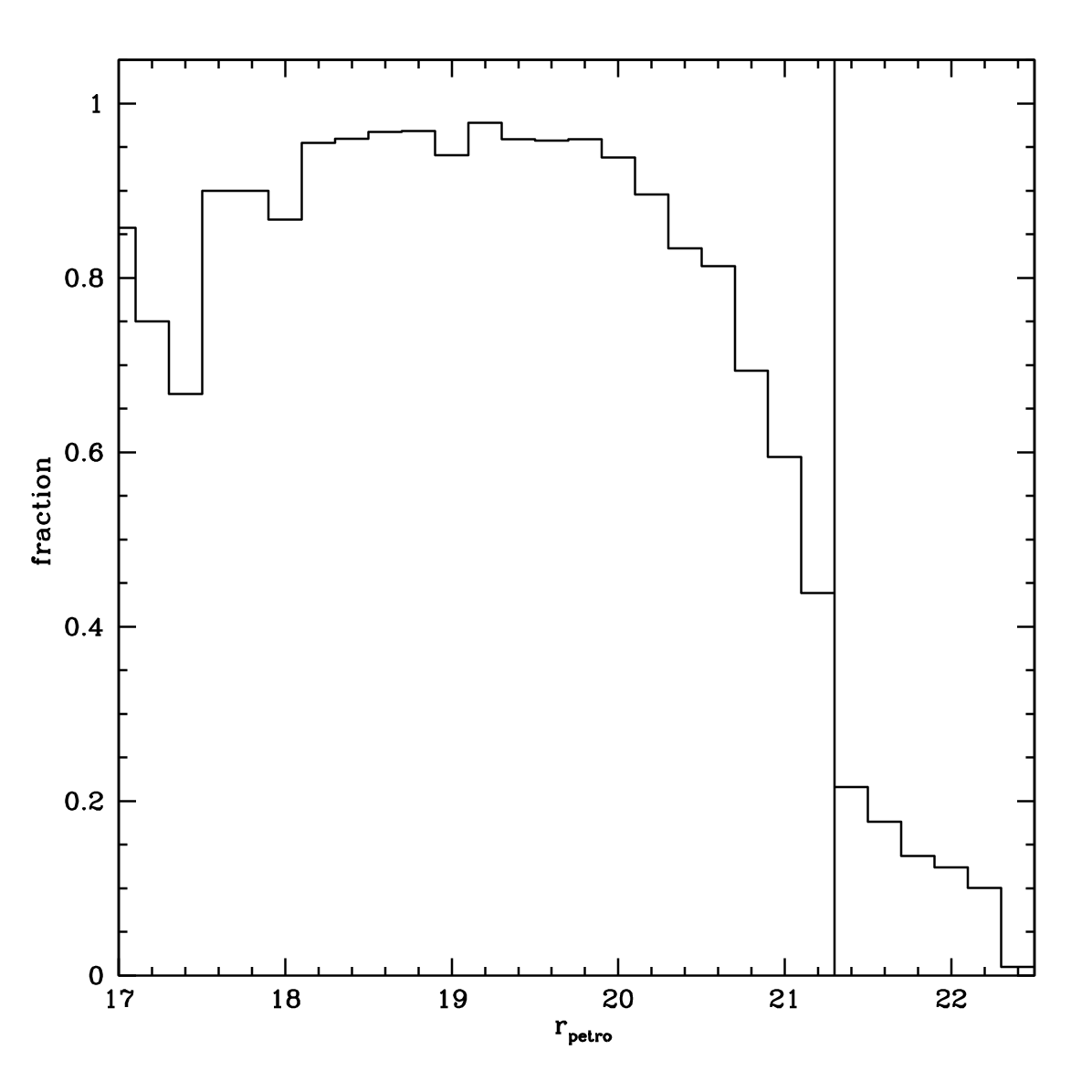}}
\vskip -5ex
\caption{GTO2deg$^2$ redshift survey completeness. The vertical line separates the primary and faint galaxy samples.  The primary sample: $17.77< r < 21.3$, 
$g -r > 1.0$ and $r-i > 0.4$. For galaxies with $r < 17.77$, we use SDSS redshifts. 
The faint sample contains galaxies with $r > 21.3$.
} 
\label{fig:completeness.subaru.histo.ps}
\end{figure}

Our first priority targets in the GTO2deg$^2$ field are galaxies with $r_{petro} \leq 21.3$,  $r-i > 0.4$ and $g-r > 1.0$. The bright limit is the SDSS completeness limit ($r_{petro}$ = 17.77); we observe only fainter galaxies for
efficiency. Our target selection yields 3992 objects. In this subsample, we obtained spectra for 3021 galaxies and 149 stars. In the original selection, we  include everything the SDSS classifies as a galaxy; spectroscopy shows that this selection includes some stars. Among the remaining 822 objects without spectroscopy, we expect $\sim 5\%$ stars. The integral completeness is thus $\sim 79\%$. 
Figure \ref{fig:completeness.subaru.histo.ps} shows that the differential completeness drops to $\sim 40\%$ at $ r = 21.3$. 

\begin{figure}
\begin{center}$
\begin{array}{c}
\includegraphics[width=5.5in]{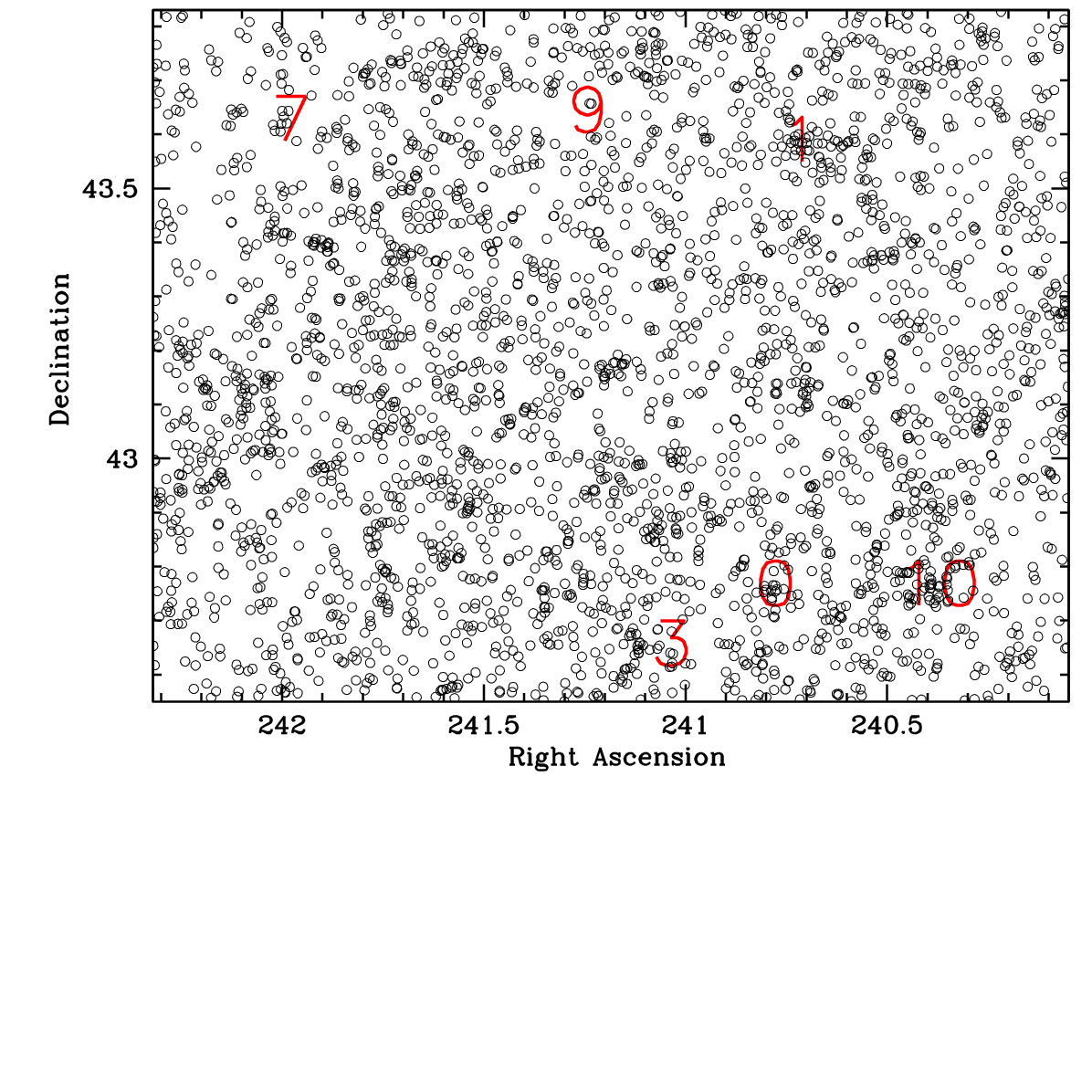} \\
\includegraphics[width=5.5in]{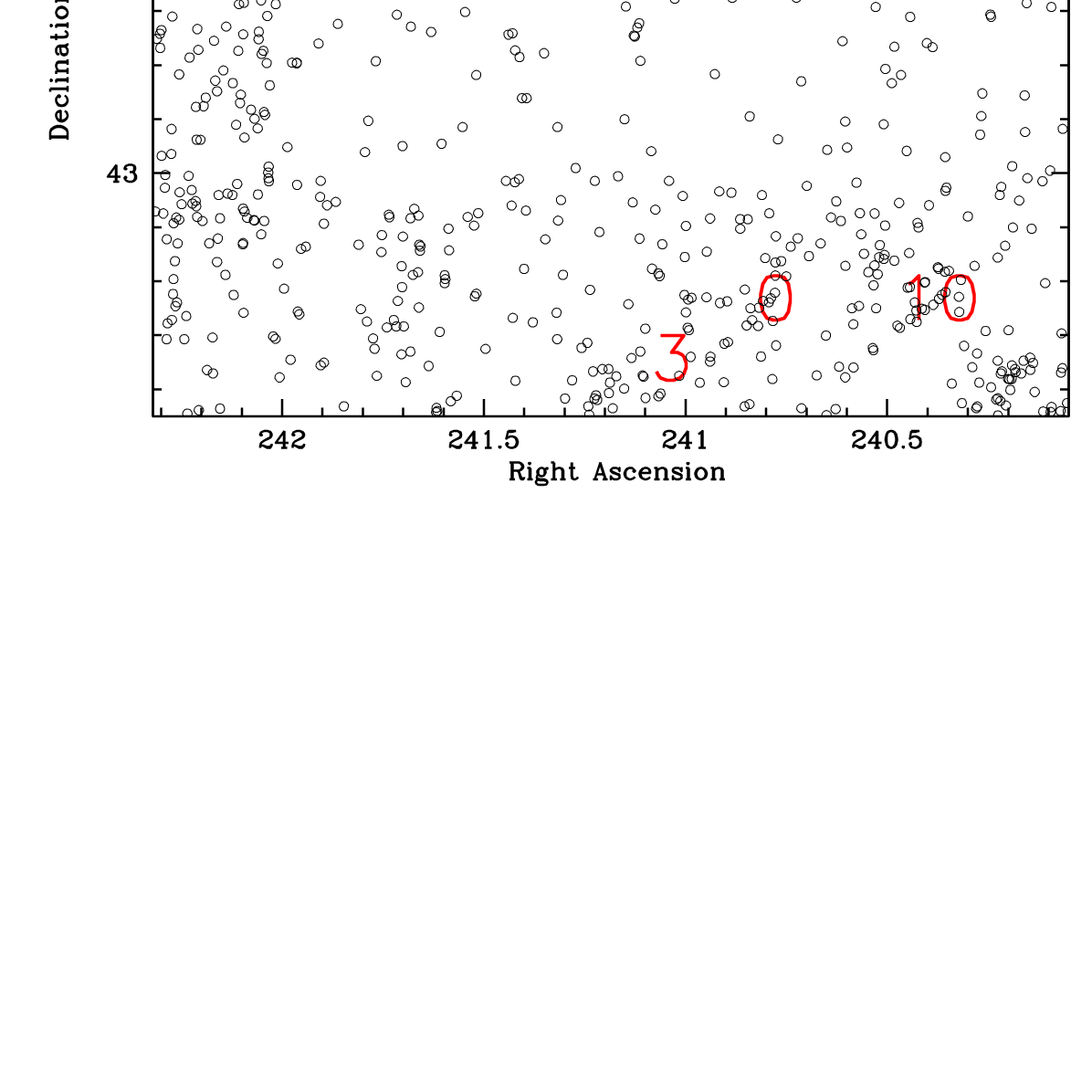}
\end{array}$
\end{center}
\vskip -45ex
\caption{Positions of the 3021 galaxies (open circles)  with a redshift in the primary sample (upper panel) and the 822 objects without 
a redshift (lower panel). The large numbers are the positions of the reliable peaks in the Subaru
weak lensing map. The completeness of the redshift survey declines toward the edges of the field.   
}
\label{fig:nozandz.ps}
\end{figure}

In Figure \ref{fig:nozandz.ps} the open circles show the positions of galaxies within the primary sample with (upper panel)  and without (lower panel) a redshift. The pattern of incompleteness is a natural consequence of the fiber positioning constraints. 
Figure \ref{fig:nozandz.ps} also shows the positions of the 6 robust convergence map peaks (see Section \ref{newmap} and Table \ref{tbl:VDisp}). Note that the numerical designations of the robust peaks are not sequential because there are peaks of intervening rank that are not secure (Section \ref{newmap}). 

\begin{figure}
\begin{center}$
\begin{array}{c}
\includegraphics[width=5.3in]{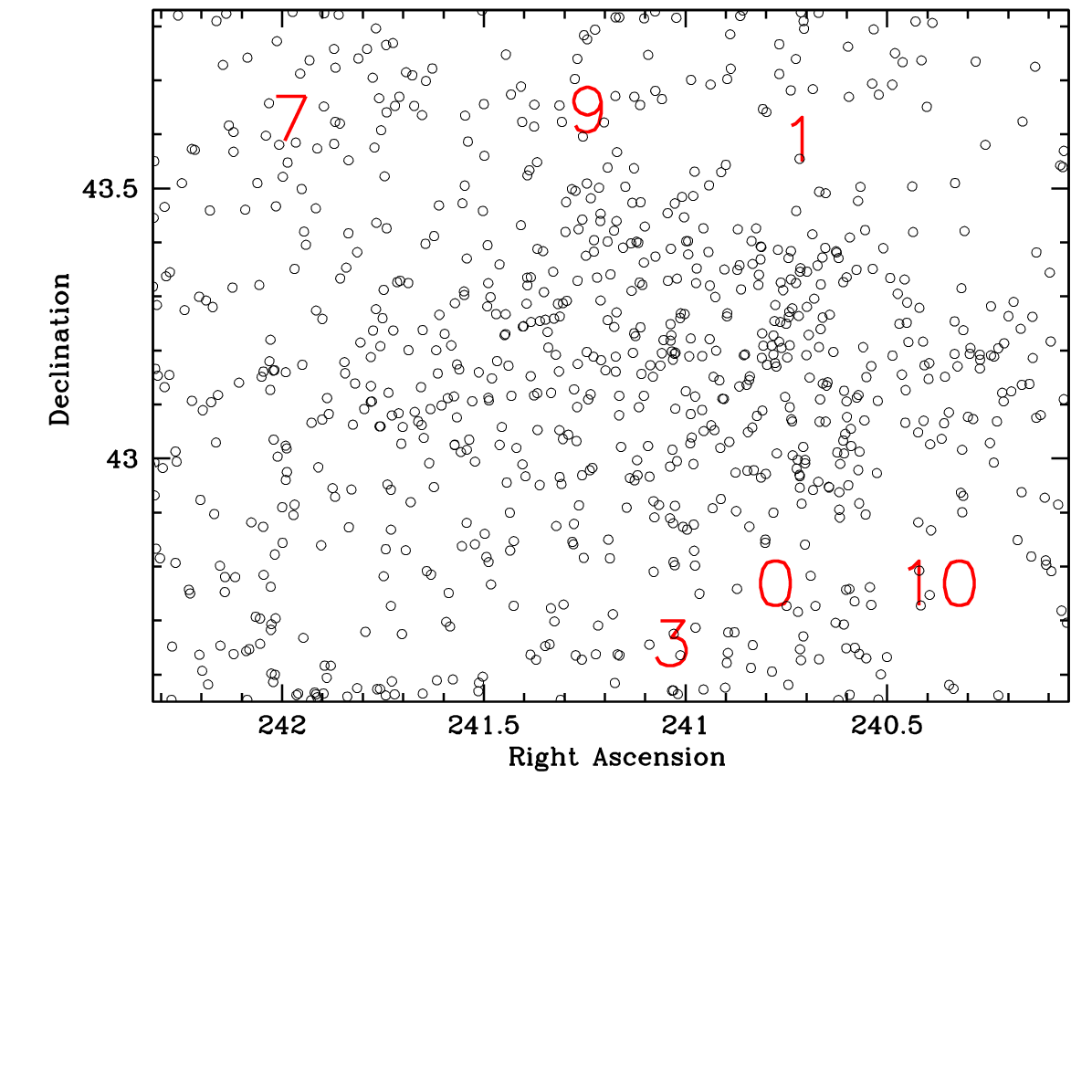} \\
\includegraphics[width=5.3in]{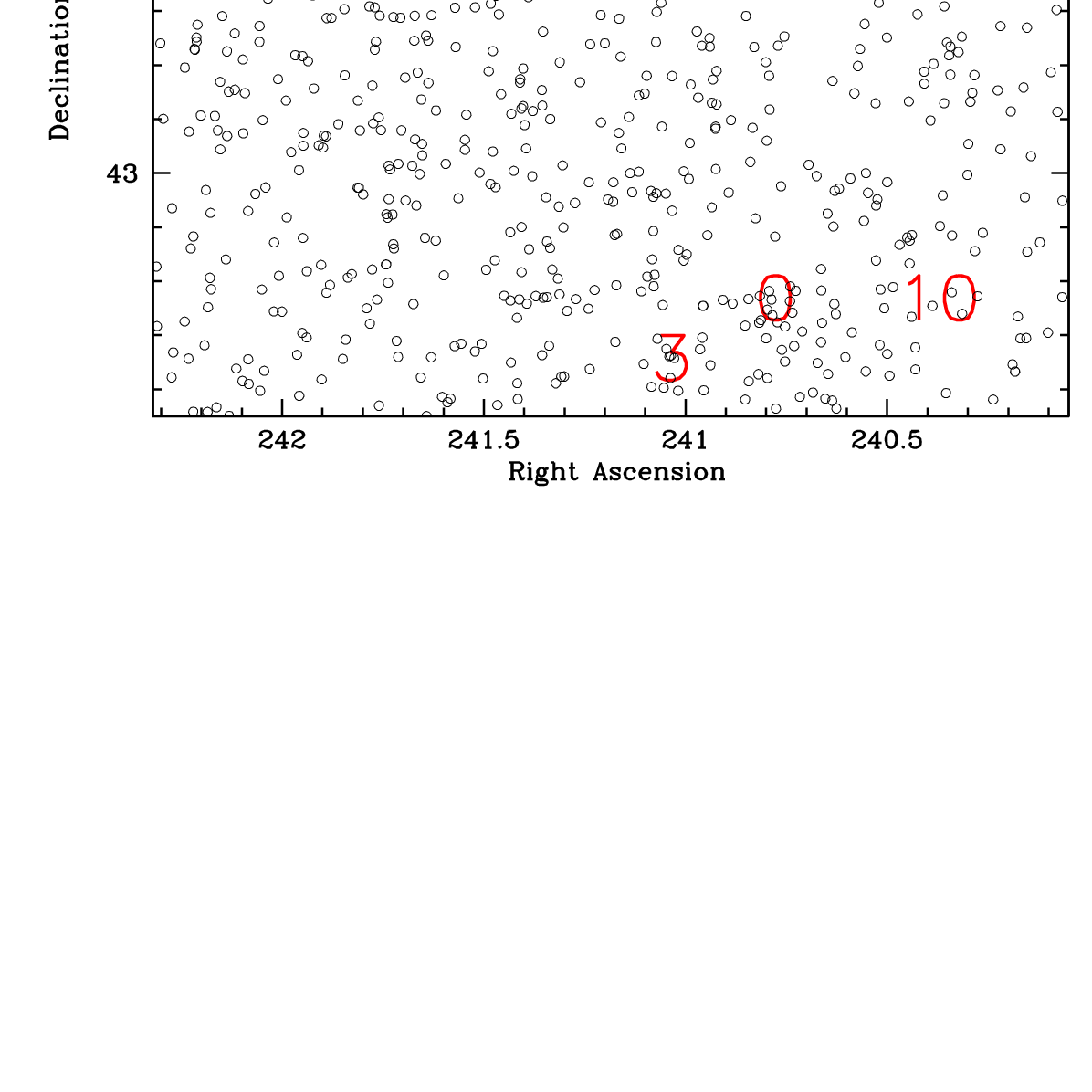}
\end{array}$
\end{center}
\vskip -45ex
\caption{Positions of galaxies with a redshift in the secondary samples. Open circles indicate
galaxy positions; red numbers indicate the reliable weak lensing peaks. The upper panel shows the distribution of the 974 bluer galaxies with $ r < 21.3$, $g-r < 1$, and
$r - i < 0.4$. Some of these objects come from the SDSS; the remaining objects fill fibers we cannot allocate to the primary targets. These objects are obviously concentrated toward the center of the field. The lower panel shows the distribution of 546 red galaxies with $r > 21.3$.
}
\label{fig:faintandblue.ps}
\end{figure}

Our secondary targets are (i) galaxies with $r > 21.3$, $r-i > 0.4$, $g-r > 1.0$ ranked by central surface brightness and (ii) bluer objects with $r < 21.3$. We observe the objects with  $r > 21.3$ beginning with the highest surface brightness objects. The lower panel of Figure \ref{fig:faintandblue.ps} shows the distribution on the sky of the  546 faint red galaxies with redshifts. The upper panel shows the distribution on the sky of the 974 bluer objects with a redshift. The faint red objects are reasonably uniform across the field; the blue objects are concentrated toward the center of the field as a result of the positioning of the Hectospec fields. Covering the edges of the field is relatively inefficient because the 1 degree diameter Hectospec field overlaps the boundary.

\begin{figure}[htb]
\centerline{\includegraphics[width=7.0in]{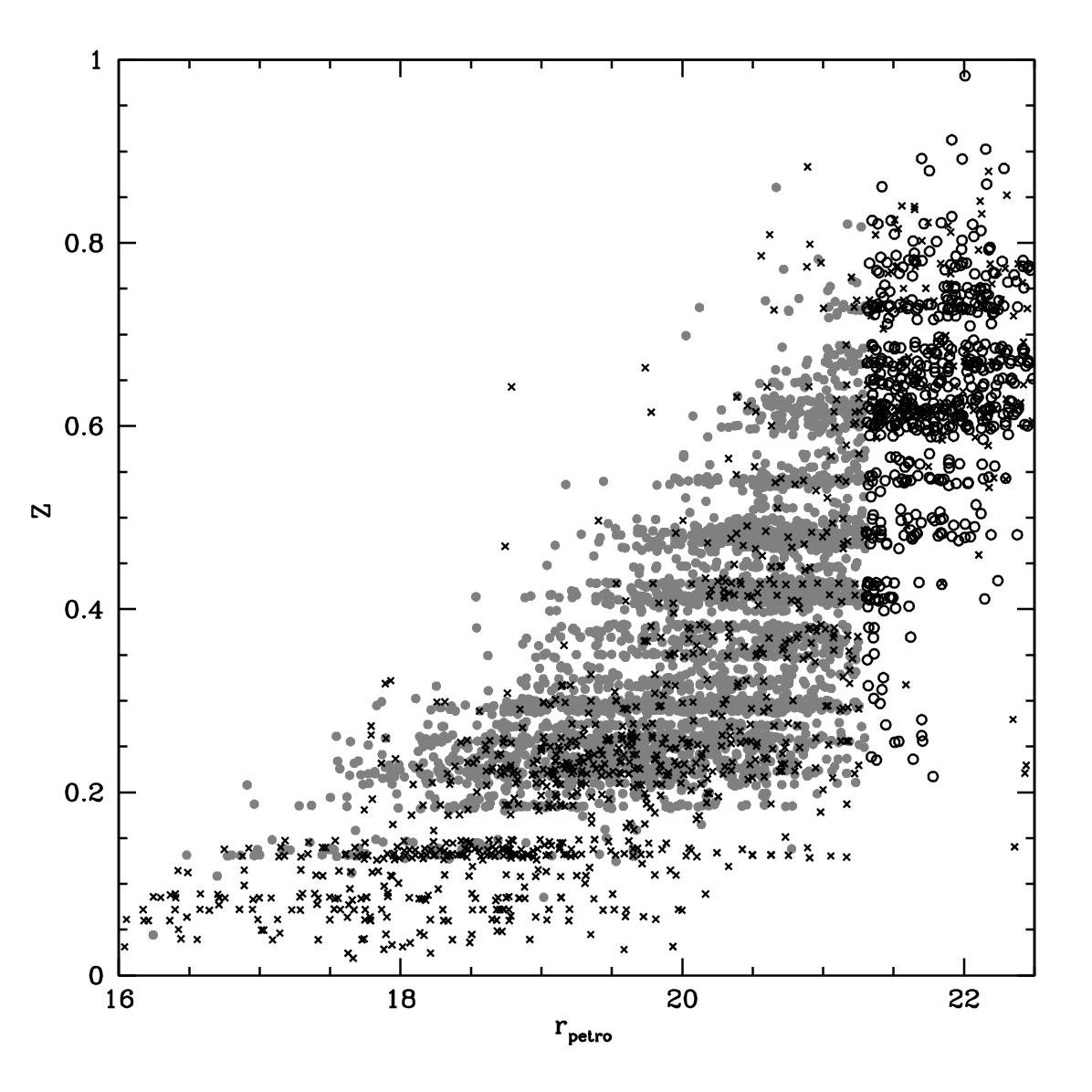}}
\vskip -5ex
\caption{Apparent magnitude as a function of redshift for the GTO2deg$^2$ redshift survey. Different symbols indicate the three subsets of galaxies plotted in Figures  \ref{fig:nozandz.ps} and \ref{fig:faintandblue.ps}. Gray dots indicate galaxies in the primary sample. Open circles indicate galaxies with $r > 21.3$ and crosses indicate bluer objects. Note that the bluer objects concentrate toward lower redshift and the fainter objects concentrate toward greater redshift as expected. 
\label{fig:rVSz.subaru.ps}}
\end{figure}

Figure \ref{fig:rVSz.subaru.ps} shows the distribution of
apparent magnitude,  $r_{petro}$, for all of the galaxies with measured redshifts. Different symbols indicate the three subsamples. At every redshift $\lesssim 0.7$ the redshift survey samples more than a one magnitude range in $r_{petro}$.

\begin{figure}[htb]
\centerline{\includegraphics[width=7.0in]{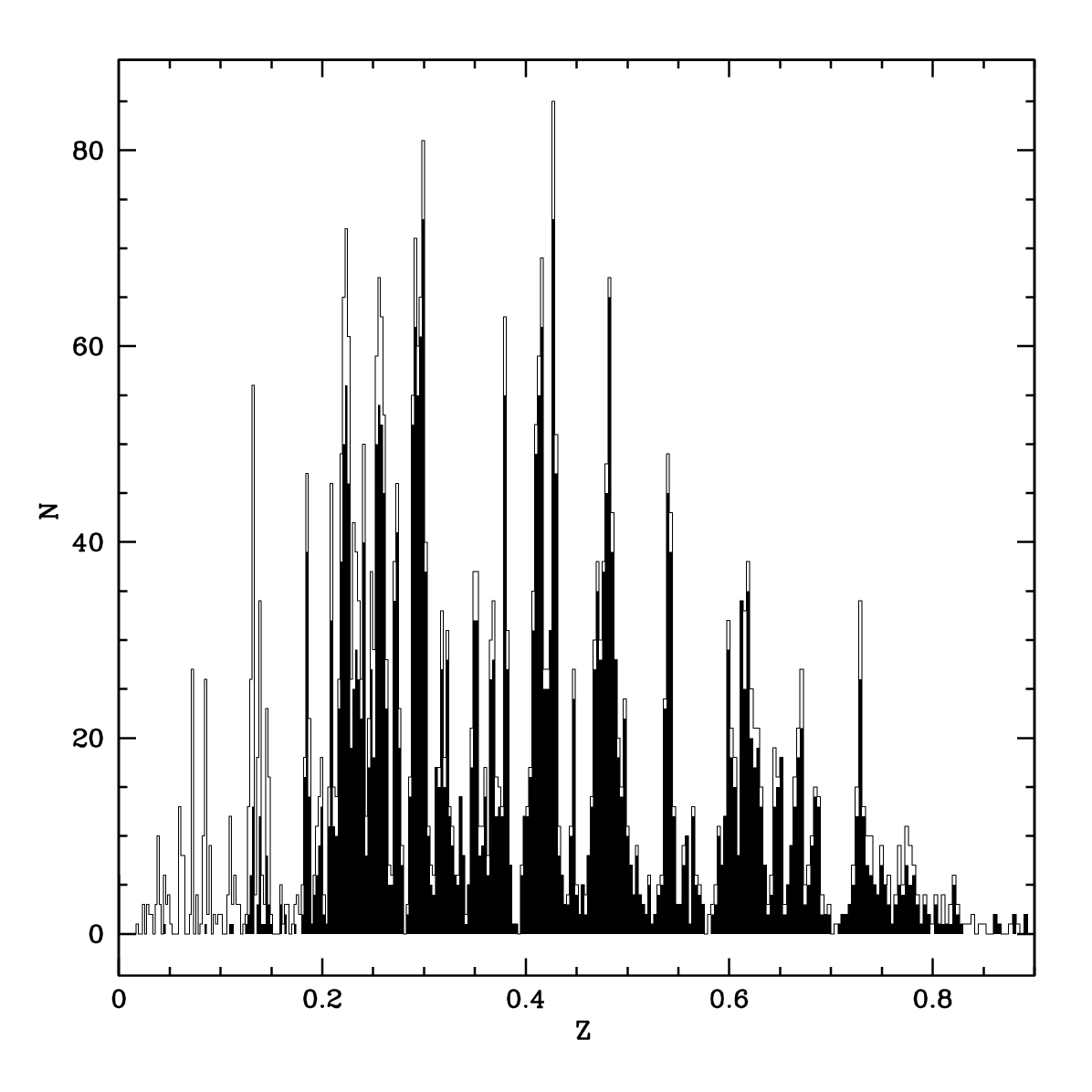}}
\vskip -5ex
\caption{Redshift histogram for the GTO2deg$^2$ survey (4541 redshifts). The bins have width $0.002(1 + z)$.The black histogram shows the redshift distribution for the primary sample of red galaxies. The open histogram shows the additional galaxies in the fainter red galaxy sample (at large redshift) and the
bluer sample (lower redshift). The narrow peaks are the standard signature of large-scale structure.
\label{fig:redshift.histogram.ps}}
\end{figure}

Figure \ref{fig:redshift.histogram.ps} shows the redshift distribution for the entire GTO2deg$^2$ field survey and for the primary target
sample. The median redshift for the primary target sample of 3025 galaxies is 0.336; for the entire sample, the median redshift is 0.349. The galaxies with $r > 21.3$ populate the high redshift range and the SDSS galaxies fill in the survey at low redshift. The sharply defined peaks are the expected signature of the large-scale structure of the universe.

\begin{figure}[htb]
\centerline{\includegraphics[width=7.0in]{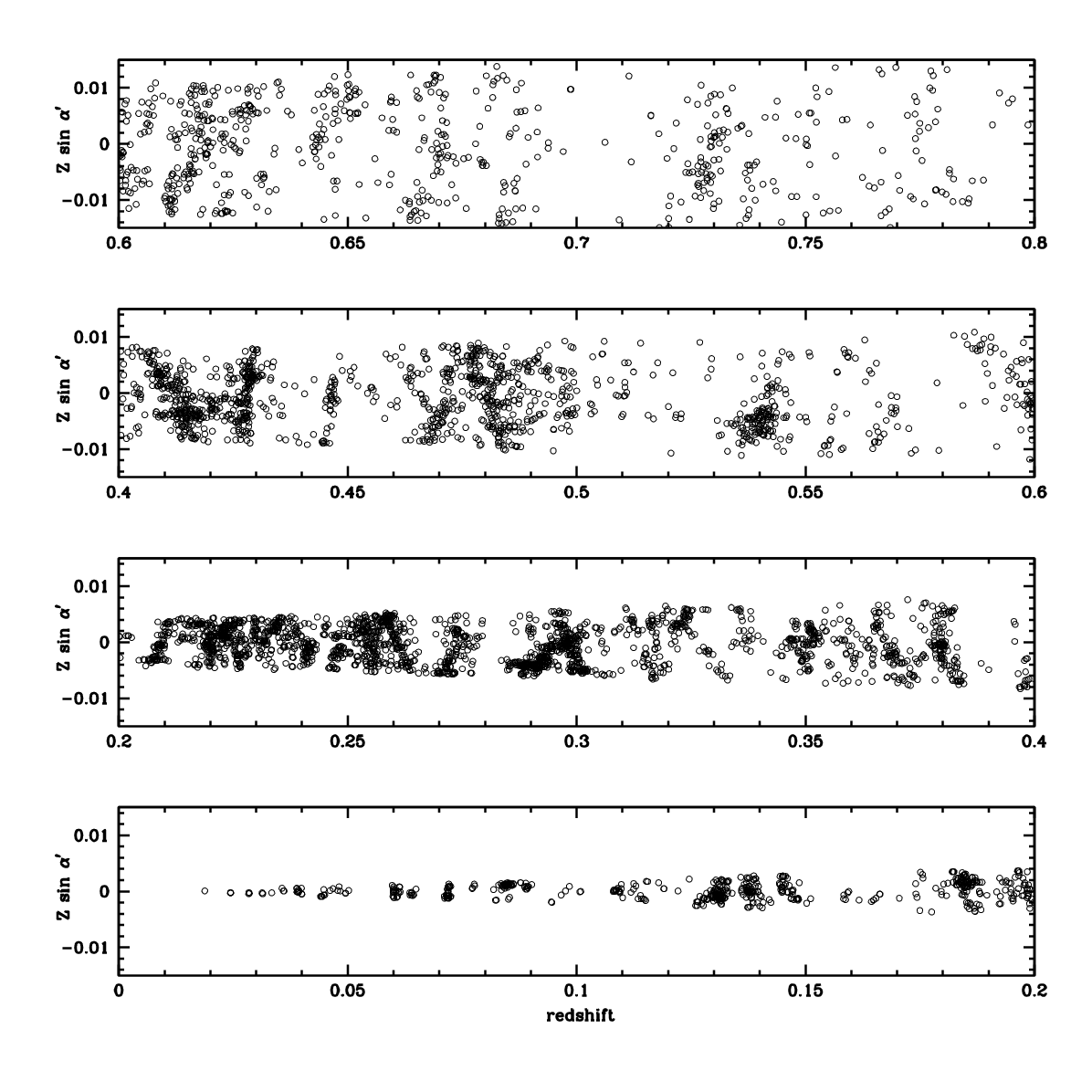}}
\vskip -5ex
\caption{Cone diagram for the GTO2deg$^2$ redshift survey in right ascension. We project the
entire declination range onto the right ascension-redshift plane.  Note the fingers corresponding to rich clusters at redshifts 0.413 and 0.540. These clusters correspond to the 
two most significant peaks in the Subaru weak lensing map (Section \ref{clusters}). The selection of primarily red galaxies enhances the contrast between the dense structures and the voids.
\label{fig:cone.diagram.subaru.ps}}
\end{figure}

Figure \ref{fig:cone.diagram.subaru.ps} shows cone diagrams for the survey. Obvious voids are delineated by walls and filaments. The contrast between the voids and the surrounding structure is enhanced relative to a redshift without color selection for red galaxies. The red galaxies are more clustered and preferentially populate more dense regions. Fingers extending along the line-of-sight  corresponding to massive clusters of galaxies are evident by eye at $z = 0.415$ and at $z = 0.540$. In Section \ref{clusters} we show that these two prominent systems correspond to the two most significant weak lensing peaks.

\section {The Weak Lensing Convergence Map}

We now proceed toward testing an updated version of the Miyazaki et al. (2007)
GTO2deg$^2$  convergence map against our dense redshift survey covering the entire field.
The redshift survey generally enables separation between massive clusters and superpositions along the line-of-sight and provides an objective test of the fraction of weak lensing peaks that correspond to massive systems. In principle, the redshift survey identifies massive systems undetected in the convergence map. 

We 
reconstruct the weak lensing map of the Miyazaki et al. (2007) GTO2deg$^2$ field to take advantage of improvements in the Suprime-Cam reduction software and to refine the procedures we used to construct the map. We adopt
mean stacking of the raw images rather than the median stacking adopted by Miyazaki et al.  (2007). This revised procedure reduces the effect of variations in seeing from one frame to another.  Otherwise we reduce the raw images as fully described in Miyazaki et al. (2007).

We describe the construction of the stacked images including steps taken to improve the PSF correction in  Section \ref{newimage}. We describe the construction of the new shear map and the B-mode systematic
tests in Section \ref{newmap}. We model the expected sensitivity of the map to massive systems of galaxies in Section \ref {sensitivity}.

\subsection {Updated Images and PSF Correction}
\label{newimage}

The GTO2deg$^2$ field consists of 9 Suprime-Cam pointings acquired on April 18-25, 2001. The exposure time varies from subfield to subfield.  When possible for a subfield, we select the best seeing images with summed exposure time of at least 2000 seconds. We use  all the exposures regardless of seeing for those fields where the total exposure time is less than 2000 seconds.  The exposure time per field for the stacked images ranges from 1800 seconds to 2400 seconds.
The median seeing for the 9 frames ranges from 0.64$^{\prime\prime}$ to 0.77$^{\prime\prime}$ with an overall median of $0.66^{\prime\prime}$.

\begin{figure}[t]
\begin{center}$
\begin{array}{cc}
\includegraphics[width=3.0in]{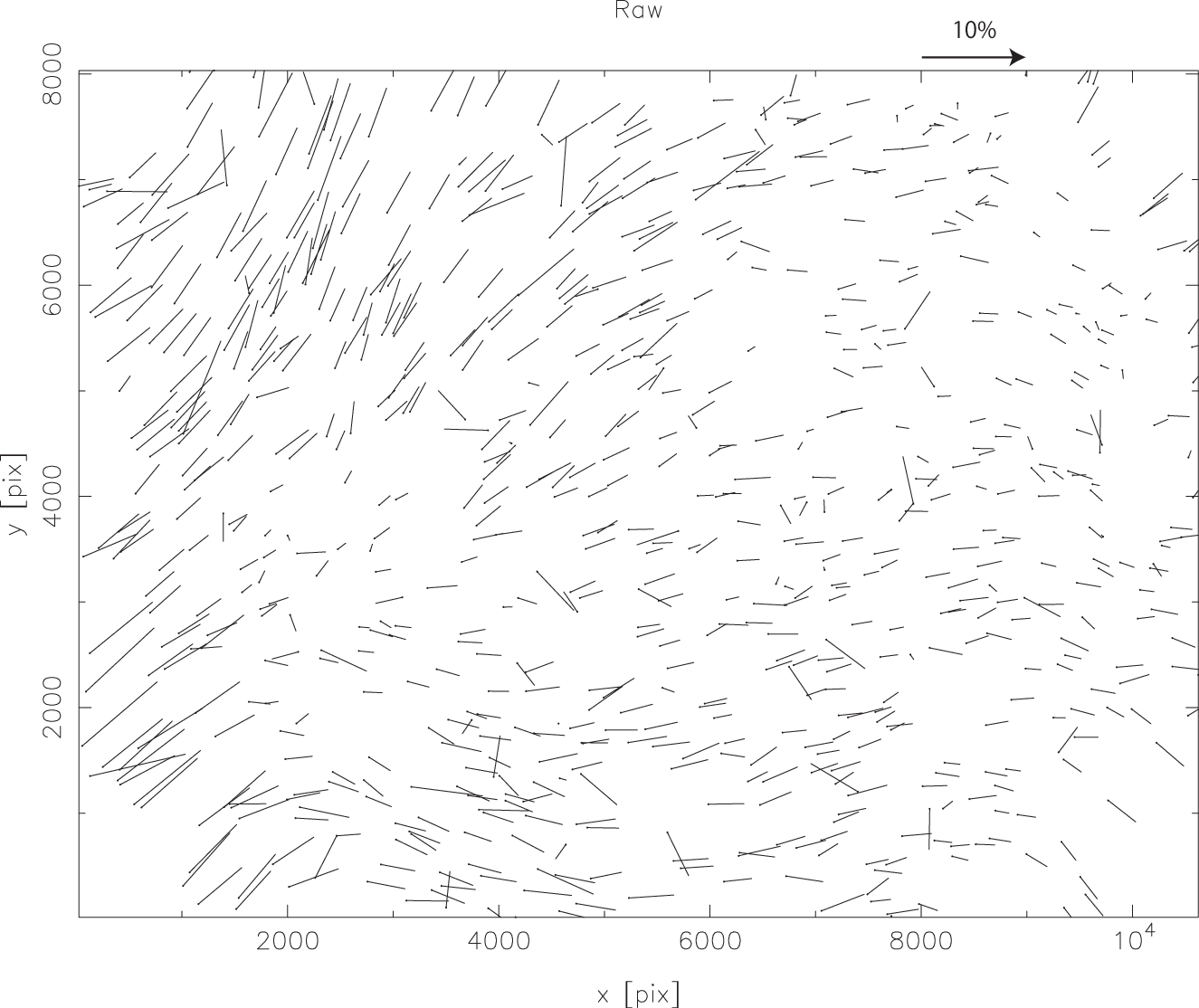} &
\includegraphics[width=3.0in]{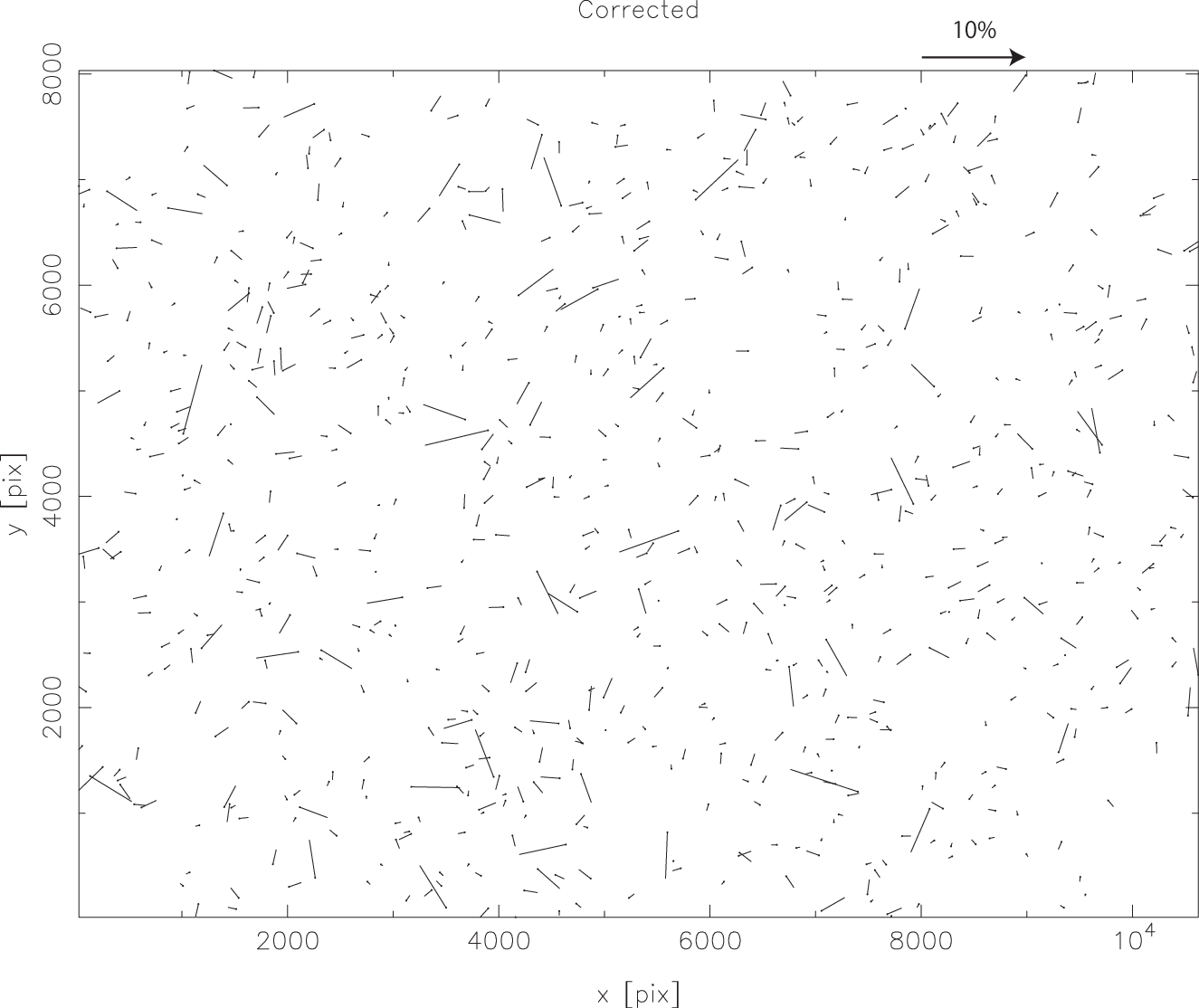}
\end{array}$
\end{center}
\vskip 7ex
\caption{Whisker plot (left panel) showing  ellipticity versus position for stars in subfield 3
of the GTO2deg$^2$ field before PSF correction.  The arrow above the diagram shows a 10\% ellipticity measurement for comparison.  A clear systematic pattern  spans the 
entire image, with maximum amplitude greater than 10\% in the left-hand corners of the image.
The right panel shows the whisker plot for subfield 3 after PSF correction.  The stars are the same as  in the left panel, but the systematic pattern is now absent and the mean ellipticity 
of stars is within measurement error in most cases.  The stars with significant ellipticity are biased by light contamination from nearby objects.}
\label{ecorrected3.eps}
\end{figure}

The reductions mainly follow the prescription given in Miyazaki et al.(2007).
The primary source of systematic error in the shear maps comes from the spatial variation in the PSFs.  Raw images from  Suprime-Cam show systematic patterns in the shapes 
of stars, primarily at the corners of the field.   We use a polynomial fit for the PSF ellipticity components of stars as a function of their position in the image (see Miyazaki etal. 2007).  Figure \ref{ecorrected3.eps} shows the whisker plot of the ellipticity of stars (selected according to their position in a size-magnitude diagram) as a function of pixel position in a 
subfield of the GTO2deg$^2$ region.

We construct an astrometric distortion solution for each chip by comparing the 
pixel coordinates of 70-100 comparison stars with their coordinates on the reference exposure.  This distortion solution takes into account the camera distortion (a radial function of  distance from the optical axis) and displacements and rotations of individual detectors compared to their fiducial positions.
Because this astrometric correction still yields residuals  too large for accurate image stacking, we calculate a residual distortion. We fit a second order polynomial to the residual distortions in object positions between the reference exposure and each subsequent exposure.  

\begin{figure}[htb]
\centerline{\includegraphics[width=5.0in, angle = -90]{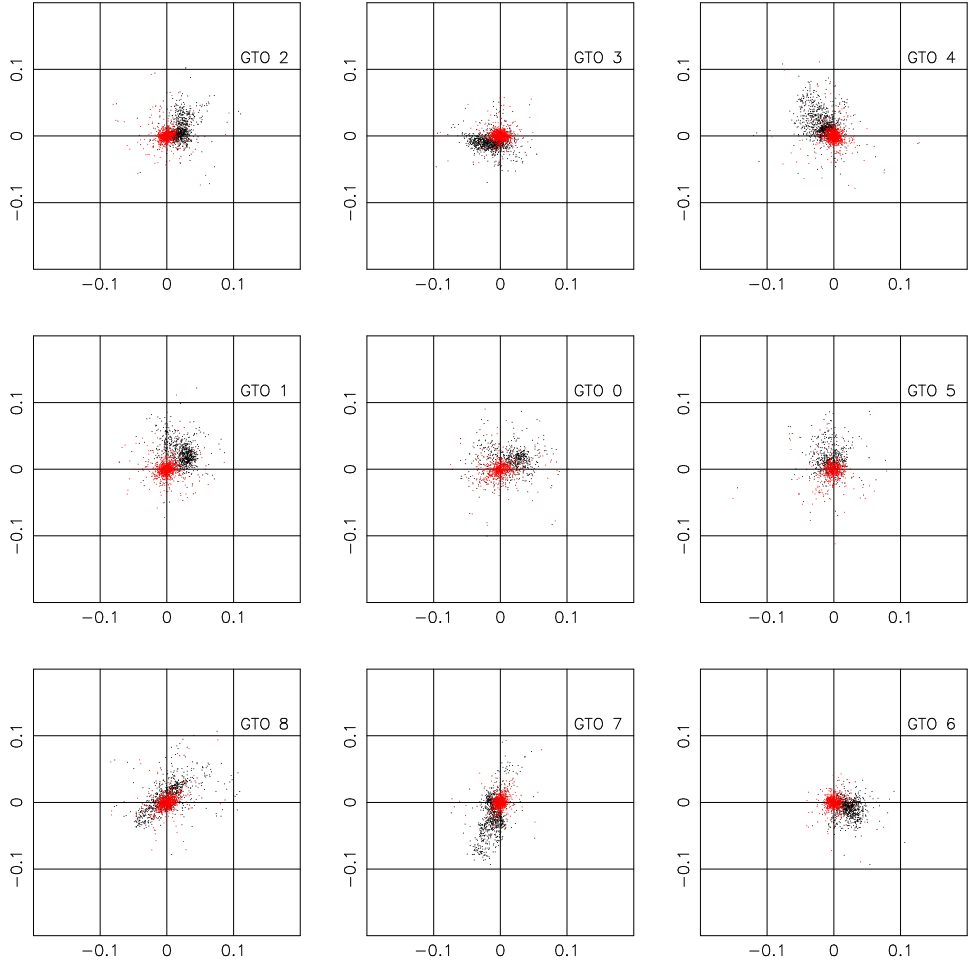}}
\vskip 7ex
\caption{ Plots of the distribution of stellar ellipticity components $e_1$ and $e_2$ before 
and after PSF correction.  Each plot represents one of the subfields of the GTO2deg$^2$ 
field, arranged accoring to their position in the final map.  Black points represent the
ellipticity components of stars measured prior to PSF correction;  red points represent
the ellipticity components for the same stars after PSF correction.  Note that the procedure completely removes the systematic shift in stellar ellipticity of both components for all subfields. Furthermore, the elliptiticy errors are isotropic in ($e_1$,$e_2$) space to within $\sim 25$\% after the PSF correction.}
\label{psf.ps}
\end{figure}

After we have corrected for the residual PSF ellipticity we obtain the residual pattern  
in Figure \ref{ecorrected3.eps}.
The systematic pattern is corrected on scales larger than the $\sim 1$ arcminute scale probed by the stars.  Another way of showing the effect of this correction is to measure the distribution of 
$e_1$,$e_2$ for stars in the subfields.  Figure \ref{psf.ps} shows the
ellipticity components of stars in the nine subfields of the GTO2deg$^2$ field, both before and after correction.  The PSF correction removes  systematic ellipticities (offsets from the origin in ($e_1$,$e_2$) space) along with most of the PSF error anisotropy (deviation from uniform scatter in all directions). The remaining PSF anisotropy error probably results from the residual PSF errors on scales smaller than clusters of galaxies.  Averaging over the larger cluster scale reduces the impact of these errors on cluster detection.  They do not contribute significant bias in cluster identifications.

We finally stack the registered images by averaging the pixel values for all unmasked 
exposures at each pixel.  The ellipticities of the stars identified on the stacked exposures are consistent with those on the individual exposures.

\subsection { An Updated Convergence Map}
\label{newmap}

To construct the convergence map, we use all galaxies with 23 $ < R_{C} <$ 26 that are detected at
significance  $ \gtrsim 10 \sigma$ as sources.
On average there are 35 sources arcmin$^{-2}$; a total of 254,010 sources meet our cuts. To derive the ellipticities of the
sources, we use the routines {\it getshapes}, {\it efit}, and {\it ecorrect} in the 
{\it imcat} package (Kaiser, Squires \& Broadhurst 1999: KSB).

From the shears, we construct lensing maps using a modified version of the 
KSB algorithm (Hoekstra et al. 1998).  First, we measure the size and ellipticity of all the stars identified in the stacked images.  We then use a second order polynomial function in pixel $x$ and $y$ coordinates to fit for the PSF size and ellipticity in each subfield image.  This procedure yields an estimate of the PSF size and shape at each galaxy position, and, therefore, the stellar shear and smear polarizabilities, $P_{sh}^*$ and $P_{sm}^*$,
respectively.

To calculate the PSF shear and smear polarizability appropriate for each galaxy's position and size, we first calculate the polarizability tensor for a set of Gaussian kernels with FWHM ranging from 0.8\arcsec\ to 4.0\arcsec\ (in increments of 0.8\arcsec).  At each position, we calculate the appropriate polarizabilities $P_{sh}$ and $P_{sm}$  by linear interpolation of the polarizability components for the kernels that bracket the galaxy size.
We can thus calculate the induced polarizability of the galaxy as (Hoekstra et al. 1998; KSB)
$$  P_\gamma = P_{sh} - P_{sm}P_{sh}^\star \left(P_{sm}^\star\right)^{-1}.$$

Because $P_\gamma$ can be affected by uncertainty in the size determination of  individual galaxies, our corrected shear measure for each galaxy is
$$ \vec{\gamma} = \vec{e}\left<P_\gamma\right>^{-1}$$

\noindent where $\vec{e}$ is the measured ellipticity and orientation of the galaxy
and $\left<P_\gamma\right>$ is the average polarizability of the 20 nearest neighbors to each galaxy we consider.

\begin{figure}[htb]
\centerline{\includegraphics[width=6.0in, angle=-90]{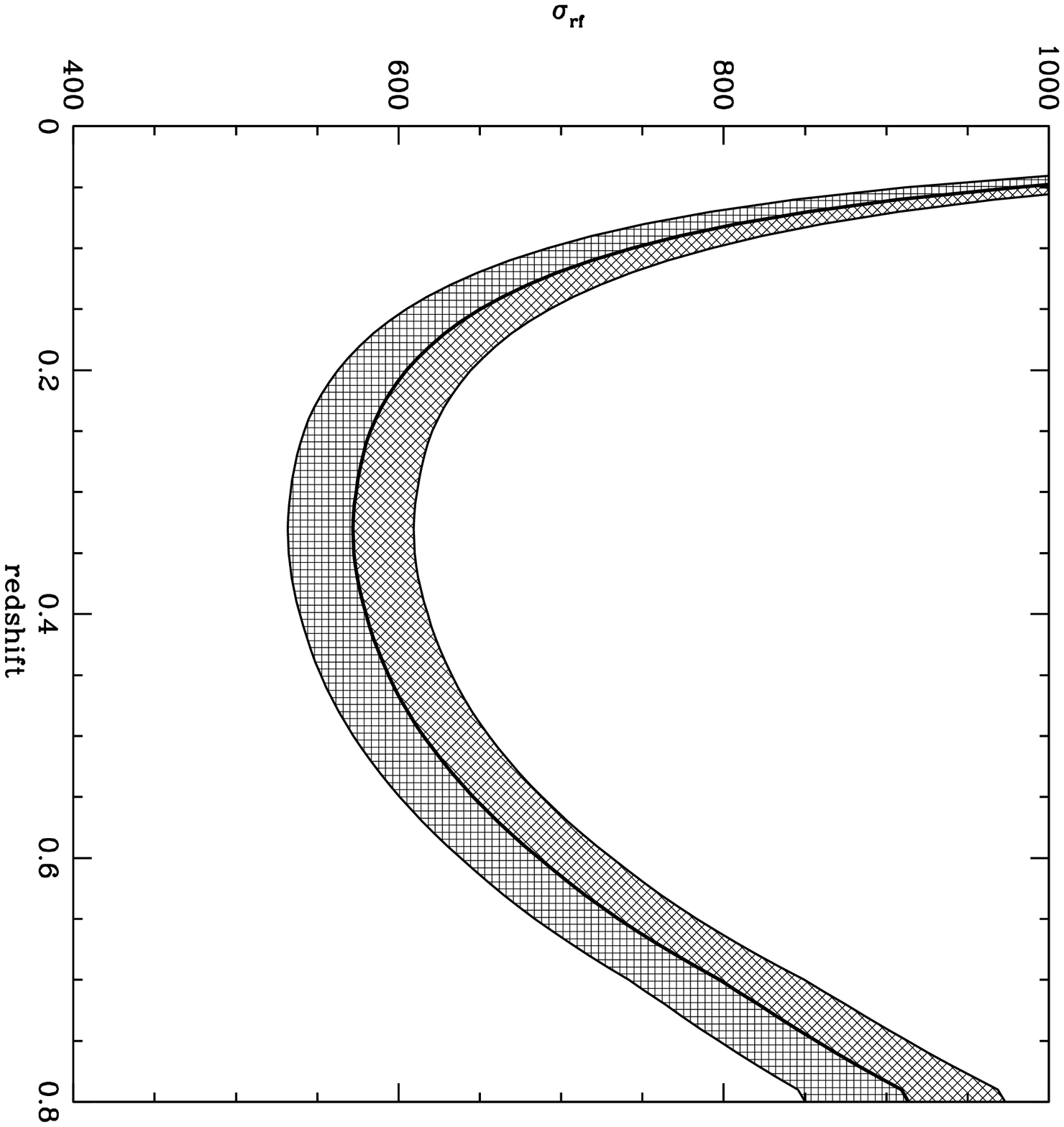}}
\vskip -1ex
\caption{GTO2deg$^2$ $\kappa$-S/N map; numbers indicate the positions of peaks with
significance $\nu > 3.7$. The lowest contour is $\nu = 2 $. The value of the number gives the rank of the convergence map peak.  Large numbers are reliable peaks; small numbers are unreliable peaks. The X marks the position of a cluster detected in the redshift survey, but it is too close to the edge of the survey for reliable detection in the lensing map; open squares are sample directions which are well-populated in the redshift survey but where there is no weak lensing peak significant at $\nu \gtrsim 2$; A denotes
the position of the Abell cluster 2158.
\label{fig:GTOkappawopeak.ps}}
\end{figure}

Once we have produced a $\kappa$ map from the shear distribution, we estimate a noise map by taking the variance over 100 realizations of the $\kappa$ map.  In each realization, we randomize the orientation of the galaxies (leaving their positions fixed) and regenerate the map. We show the $\kappa$-S/N map (the $\kappa$ map divided by the variance map) in 
Figure \ref{fig:GTOkappawopeak.ps}.

We next use Sextractor to identify peaks in the $\kappa$-S/N map. The pixels in the map are
$0.3^{\prime} \times 0.3^{\prime}$, and we use a smoothing kernel with a $1.5^{\prime}$ FWHM. 
A peak must have at least 3 connected pixels significant at the 2$\sigma_{\kappa-S/N}$ level or more to be selected; $\sigma_{\kappa -S/N}$ is the variance in the $\kappa$ -S/N map.

The numbers in Figure \ref{fig:GTOkappawopeak.ps} are the significance rank of  the 11 peaks we find. Table \ref{tbl:VDisp} lists the significance of all of these 11 peaks with  
signal-to-noise greater than 3.7$\sigma_{\kappa-S/N}$ or, equivalently with $\nu > 3.7$. As in Miyazaki et al. (2007) we remove peaks within 4$^\prime$ of  bright stars (USNO B1.0 R$_1 < 11$ or B$_1 < 11$) or within
2.7$^\prime$ of the edge of the field; we remove peaks 2, 4, 5, 6
and 8 for these reasons. Interestingly, we show below (Section \ref{clusters}) that these unreliable peaks do not correspond to any systems in the redshift survey.

Although we have demonstrated that our PSF correction algorithm successfully removes the errors introduced into the PSF at the position of the stars, there is still the possibility that systematic errors affect the shape determinations at the positions where the shears are being measured (i.e. at the positions of the galaxies).  To test the possibility of remaining bias in the PSF, it is customary to make shear maps using the 45-degree-rotated component rather than the tangential component of the ellipticity tensor, or, equivalently, by rotating all galaxy orientations by 45 degrees.  This ``B-mode'' map is important because weak lensing shears do not induce any signal in the 45-degree-rotated ellipticity components.  On the other hand, systematic errors in shape measurements should affect both components of ellipticity equally.  As a result, the level of signal in the B-mode map is a good indication of the level of spurious signal in the shear or (E-mode) map.
We use the noise map discussed above to normalize the B-mode map. Because the shuffling of the shape assignments randomly mixes ellipticity components, use of the same noise map is an appropriate normalization for both the convergence and B-mode maps.

\begin{figure}[htb]
\centerline{\includegraphics[width=5.0in, angle = -90]{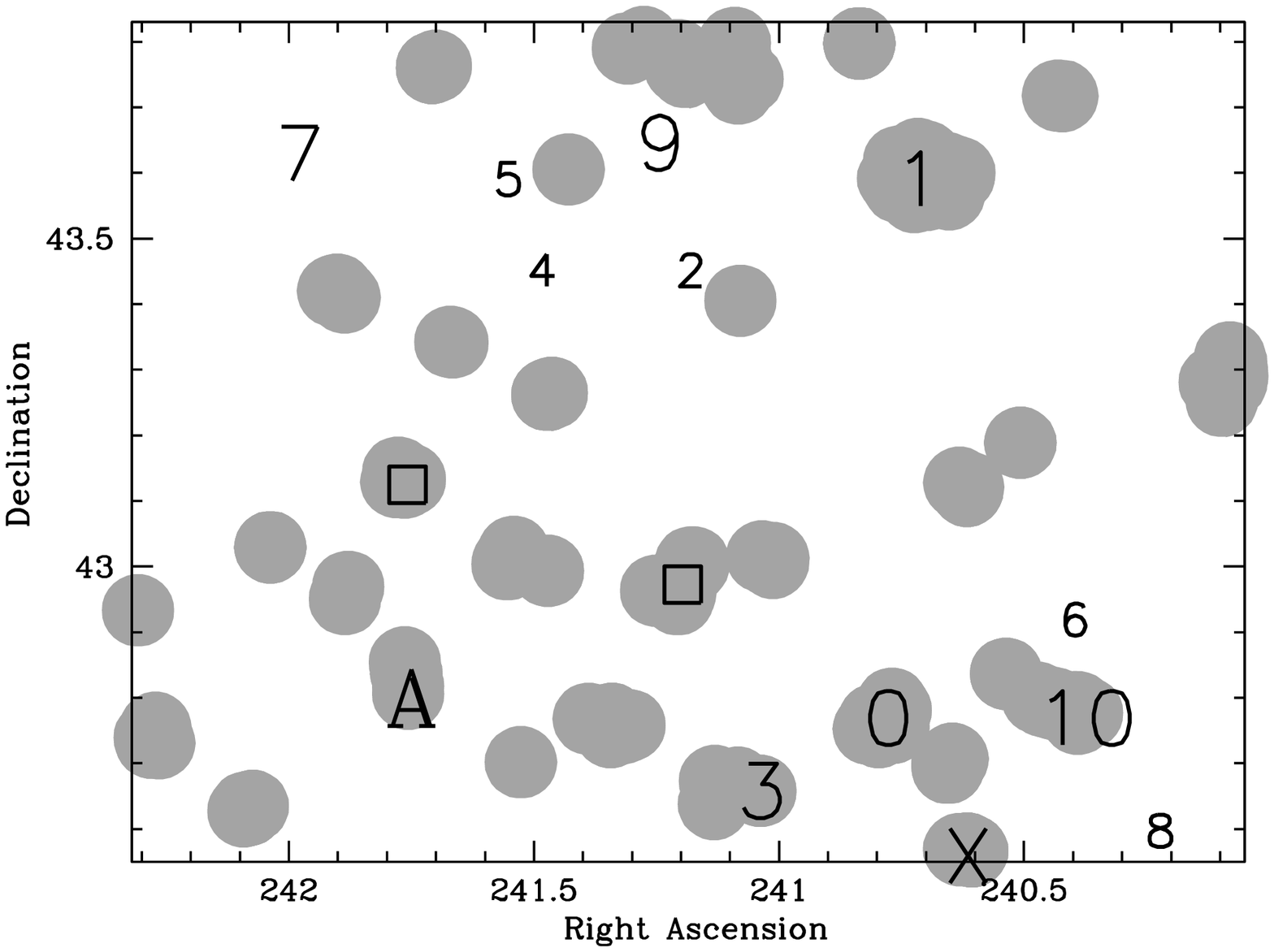}}
\vskip 7ex
\caption{``B-mode'' map of the GTO2deg$2$ field constructed using the same 
galaxies used for the shear peaks in Figure \ref{fig:sigmamap.scaled.ps}. Galaxy positions are preserved, but the shapes are rotated by 45 degrees before the measurement of the shear pattern.  Contours in the map mark $1\sigma$ intervals; the lowest contour is 2$\sigma$.  The labels correspond to the labels in the $\kappa$-S/N map (Figure \ref{fig:GTOkappawopeak.ps}).}
\label{GTO.cfa.bmode.ps}
\end{figure}

In Figure \ref{GTO.cfa.bmode.ps} we show the B-mode map for the GTO2deg$^2$ field, with signal-to-noise contours constructed in the same way as for the $\kappa$-S/N map (Figure \ref{fig:GTOkappawopeak.ps}). We use the same peak-finding algorithm as we do on the $\kappa$-S/N map to uncover peaks in the B-mode map.  The highest peak in the B-mode map has a signal-to-noise level of $\sim 4.25$. We list the coordinates and significance of the four highest B-mode peaks in Table \ref{Bmodetab}; these peaks have $3.7 < \nu \leq 4.25$.  It is interesting that Shan et al. (2011) find a similar number of high significance peaks per unit area in this significance range  in their B-mode maps derived from CFHT data (see their Figures 14 and 15). One of  the high significance B-mode map peaks is obviously contaminated by a nearby bright star (Table \ref{Bmodetab}).

The B-mode map suggests that we set our threshold  for reliable cluster
detection at $\nu = 4.25$ rather than the $\nu = 3.7$ of Miyazaki et al. (2007).  
We consider peaks in the $\kappa$-S/N map with $\nu \geq 3.7$ as an independent test of the implications of the B-mode map. In essence, Shan et al. (2011) make a similar test by examining the purity of the sample of peaks in their E-mode map as a function of $\nu$.

\subsection {Modeling the Sensitivity of the Convergence Map}
\label{sensitivity}

As an additional basis for examining the completeness and  purity of the weak-lensing selected cluster
catalog derived from the $\kappa$ map, we estimate the expected sensitivity limits of the convergence map as
a function of redshift. For any individual background galaxy in the weak
lensing limit, the induced {\it measured} tangential ellipticity (and hence the
S/N for detection) depends on both the angular diameter distance to the
galaxy and on the size of the galaxy relative to the PSF.
For a collection of galaxies, the effective total weight as a function of
redshift is a polarizability-weighted sum over the individual distance ratios.  The
effective distance ratio for the ensemble of background galaxies is

\begin{equation}
 W_{eff}(z_l) = {{\sum_{s}  \frac{D_A(0,z_l) D_A(z_l,z_s)}{D_A(0,z_s)}} W_s
\over {\sum_{s} W_s }}
\end{equation}
\\
\noindent where the subscripts $s$ and $l$ refer to the sources and lens,
respectively. $W_s$ is the weight of each source galaxy (which depends on
the object size), $z_l$ is the lens redshift,  $D_A(z_1,z_2)$ is the angular
diameter distance between $z_1$ and $z_2$, (the ratio is zero if $z_s<
z_l$), and the sum is over all sources.  For a realistic survey  an additional suppression results from misidentification of faint foreground galaxies, each of which should have weight zero. 

One of the advantages of Subaru data we use here compared to other deep ground-based surveys comes from the improved seeing.  In a typical deep ground-based image where the stellar PSF is $>0.75$\arcsec, up to half the galaxies detected at $10\sigma$ significance are unresolved (with size less than $1.2D_{PSF}$ (Kubo et al. 2009 )).   The images of the GTO field were taken in better seeing (median seeing 0.66\arcsec).  With this image quality, $\sim 80$\% of the galaxies at the object depth cutoff are resolved from stars.

For each redshift, we compute the angular diameter
distance factors in equation (1) assuming the ``concordance"
cosmology (Spergel et al. 2007). We derive the $W_s$ terms  using data from the COSMOS Subaru galaxy catalog (Taniguchi et al. 2007) and the COSMOS ACS catalog (Leauthaud
et al. 2007).  The COSMOS combined ACS/Subaru dataset is currently
the best dataset for this calibration because it covers by far the
largest area with both ground- and space-based resolution to faint limiting
magnitude.  In addition, the Subaru multi-color observations provide
photometric redshift estimates  (Mobasher et al. 2007)
for the majority of galaxies in the survey area. We use the COSMOS catalog
size and photometric redshift information to estimate the redshift
distribution of the galaxies that would be resolved in our weak lensing map.  

We match galaxies in the COSMOS Subaru and ACS catalogs with the additional requirement that the catalog I magnitudes match to within 0.3 (This restriction eliminates errors in matching the two catalogs; it also eliminates most objects that
cannot be adequately separated from neighboring objects in the Subaru
imaging).  Because the Subaru images of the COSMOS field were taken in comparable seeing (median seeing $\sim$0.55\arcsec), 
this procedure should not alter the COSMOS sample compared with the Subaru sample. The final catalog contains approximately 124,000 objects with magnitudes, sizes, and photometric redshift estimates.

The galaxy catalog of the GTO2deg$^2$ field  has sizes and magnitudes (R$_C$ magnitudes; Miyazaki et al. 2007), but not photometric redshifts.  To assign redshifts to the galaxies, we assign to every galaxy in the GTO2deg$^2$ field the redshift of a randomly selected galaxy in the COSMOS catalog with the same magnitude and (seeing-deconvolved size), where the list of of galaxies is taken from all the galaxies that match to within 0.05\arcsec\ in size and within 0.1 magnitudes.  Although this procedure does not necessarily assign the correct redshift to any single galaxy, it assigns the correct distribution of galaxy redshifts given the size and magnitude distribution of the galaxies in the GTO2deg$^2$ field, and thus a correct effective total weight via equation (1).  
Given this catalog of redshifts (which by construction has the same number of objects and the same size distribution as the one used in the construction of the lensing map), we calculate the effective weight $W_{eff}$ that would be measured as a function of lens redshift.  

We normalize the redshift dependence of our sensitivity by comparing to simulations.
To minimize the calculations, we normalize to the sensitivity at a fixed cluster redshift of 0.3, and use the calculation of $W_{eff}$ to calculate the redshift-dependence of the sensitivity curve. We follow a modified version of the procedure for generating simulated catalogs in Khiabanian \& Dell'Antonio (2008).  We generate simulated galaxies with the same size-magnitude and magnitude-redshift relations as the HDF North and South fields (Williams et al. 1996; Casertano et al. 2000) using the prescription of Khiabanian \& Dell'Antonio (2008). We use these simulated galaxies to populate images (0.04\arcsec\ resolution) representing seven logarithmically-spaced redshift shells. We then distort the
images with a lens modeled as a core-softened cutoff isothermal sphere with
a given rest frame line-of-sight velocity dispersion, $\sigma_{iso}$, a
1$^{\prime\prime}$ core radius, and a 100$^{\prime\prime}$ cutoff radius.

We resample the distorted images with the Subaru pixel scale (0.2\arcsec/pix), coadd them, and convolve with a Gaussian PSF with a 0.66\arcsec\ FWHM and add noise to match the expected noise in the Subaru GTO observations.
We repeat this procedure  for different values of $\sigma_{iso}$ (at a fixed
$z_{lens}=0.3$) to generate maps of the weak lensing signal.  We run the
Kubo et al. (2009) detection method  on these images to determine the value
of $\sigma_{iso}$ where the peak S/N reaches our $\nu = 4.25$ cluster selection limit.
In Figure \ref{fig:sensitivity.ps}, we plot the detection limit as a function
of redshift ($L(z) = L(z=0.3) {W_{eff}(z=0.3)/W_{eff}(z)}$); the solid line represents the $\nu = 4.25$ cutoff in the lensing peak selection sample.

Our method of calculating the sensitivity fails at low redshift $z<0.1$, because the simulation method of Khiabanian \& Dell'Antonio (2008) underpredicts the significance of clusters that occupy a large angular region.  However, because none of the GTO2deg$^2$ field clusters have $z<0.2$, the comparison of sensitivities is valid.

There are some fundamental limitations to the accuracy of the calculations. First, 
lensing is sensitive to structures projected along the line-of-sight (Hoekstra et al. 2001; White et al 2002; de Putter \& White 2005;  Hoekstra et al. 2011). The increase in the scatter of the shear measurements is 
$\sim 15-20\%$ for a generic cluster where the distribution of the large-scale structure around the cluster is unknown. For the purposes of our study, this effect may boost marginal peaks above 
the $\nu = 4.25$ threshold. We do not use the significance of the peaks to measure the mass and 
the redshift survey directly reveals a subset of the intervening structures.
 
As a result of cosmic variance,  the actual number and redshift distribution of the background galaxies behind any cluster may not be well-represented by our model.  For clusters near the 
$\nu = 4.25$ detection limit, the shear signal is based on the shapes of only $\sim 2000$ galaxies.  On the scale of this small field, non-uniformities in the distribution of the background galaxies are important.  For larger samples, Hoekstra et al (2011) show that this issue is less important. Our detection sensitivities should be viewed as averages of the sensitivity across the field rather than the exact sensitivity to any one cluster at any particular position in the field.

Although the COSMOS/Subaru field is extremely valuable, it covers $\lesssim 1$ square degree of the sky.  Therefore, any systematic differences between the properties
of the COSMOS field and GTO2deg$^2$ can change the normalization.  One of the results of the analysis of the DLS (Geller et al. (2010)) and the GTO2deg$^2$ field (this work) is that there are large field-to-field variations in the types and amount of structure present at all measured redshifts even on degree-sized angular scales.  Thus, although the COSMOS field is currently the best field for calibrating weak lensing detections, larger fields will be necessary.

\section {Probing the GTO2deg$^2$ Field Redshift Survey}
\label{probes}

To evaluate the correspondence between systems in the redshift survey and 
significant peaks in the convergence map, we follow the approach of Geller et al. (2010) for sampling
the redshift survey. We examine the redshift distribution in cones with radii of 3$^\prime$ and
6$^{\prime}$. The 3$^\prime$ sampling  is similar to probes used in previous assessments of the purity of weak lensing based on counts (Schirmer et al. 2007), photometric redshifts (Gavazzi \& Soucail 2007; Shan et al. 2011) and spectroscopic redshifts (Geller et al. 2010). The combination of 3$^{\prime}$ and 6$^{\prime}$ probes samples the
virial radii of massive clusters throughout the  redshift range where we might expect the convergence map to detect these systems.

The 3$^{\prime}$ radius corresponds to 0.45 Mpc for $z = 0.15$ and 
to 1.1  Mpc for $z = 0.53$. These radii are within r$_{200}$, the radius where the enclosed average mass density, $\rho(<r)_{200} = 200\rho_c$. Here $\rho_c$ is the critical density. The radius $r_{200}$, a proxy for the virial radius, ranges from 1.0 to 2.0 Mpc for well-sampled clusters with
rest frame line-of-sight velocity dispersions in the range 500 km s$^{-1}$ to 1000 km s$^{-1}$ (Rines et al. 2003; Rines \& Diaferio (2006)). According to the scaling relations of Rines \& Diaferio (2006), the corresponding range of masses, M$_{200}$, within r$_{200}$ is 1.3 $\times 10^{14}$ M$_\odot$ to 1.2$\times 10^{15}$ M$_\odot$. 

We assess the {\it completeness} of the set of massive clusters among the high confidence weak lensing candidate systems.
We define {\it completeness} as the fraction of all clusters in the foreground redshift survey with
line-of-sight velocity dispersions above the threshold curve that also correspond to weak lensing peaks
with $\nu > 4.25$. In other words, the completeness is the fraction of appropriately massive individual systems in the redshift survey that are detected as weak lensing peaks. 

We also estimate the {\it purity} of the convergence map in the identification of massive cluster halos. Here we define {\it purity} as the
fraction of convergence map peaks with $\nu > 4.25$ that correspond to an individual massive cluster identified within the redshift survey with a line-of-sight velocity dispersion above the solid $\nu=4.25$ threshold curve of Figure \ref{fig:sensitivity.ps}. In other words, the purity is the fraction of weak lensing peaks that correspond to appropriately
massive individual systems.

The definitions of {\it purity} and {\it completeness} exclude superpositions of small
groups from consideration. These superpositions may indeed produce a weak lensing signal, but the weak lensing peak should obviously not be counted as detection of a single massive system in a mass selected cluster catalog.  The definitions also exclude coincidences between the abundant low velocity dispersion systems in the redshift survey with convergence peaks. Our definitions are similar to those adopted by Hamana et al. (2004) in their simulations of the purity and completeness of weak lensing surveys for massive cluster identification.

\subsection {Evaluating Convergence Map Peaks with a Foreground Redshift Survey}
\label{sampling}

Schirmer et al. (2007), Gavazzi \& Soucail (2007) and Shan et al. (2011) use uniform
procedures for evaluating the coincidence of concentrations of galaxies 
with convergence map peaks. Their approaches are based on complete photometric surveys of their fields. They evaluate 
the significance of cluster candidates internal to their surveys. Geller et al. (2010) develop an analogous approach for evaluating the coincidence of concentrations of galaxies in a 
foreground redshift survey with convergence map peaks. Here we follow the Geller et al. (2010) approach.  

\begin{figure}[htb]

\centerline{\includegraphics[width=7.0in]{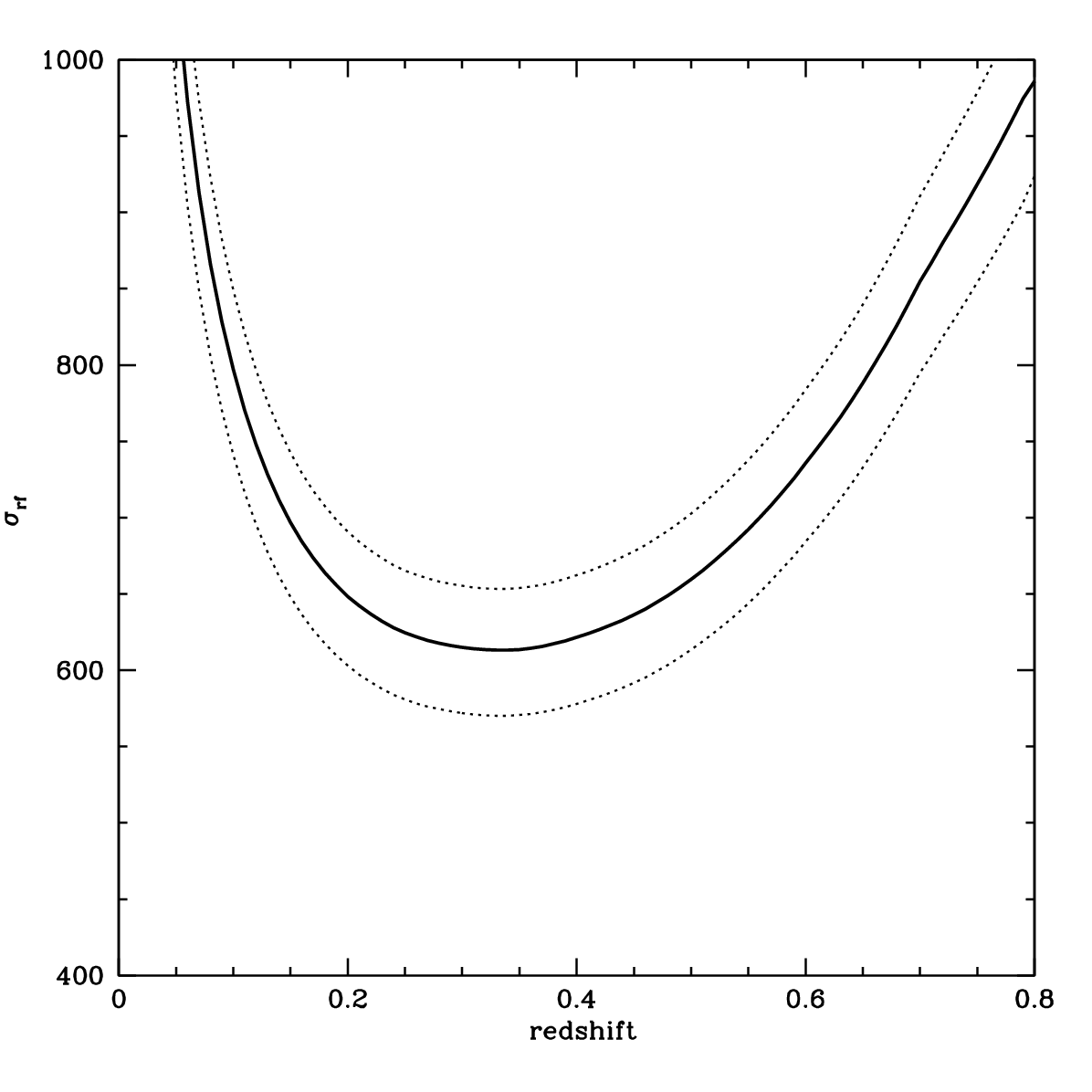}}
\vskip -5ex
\caption{Line-of-sight velocity dispersion of clusters that produce a $\nu = 4.25$ signal (central solid curve) in the weak lensing map as a function of the cluster redshift.
The shaded band is 1$\sigma_{\kappa-S/N}$ wide.
\label{fig:sensitivity.ps}}
\end{figure}

Because the redshift survey is substantially complete we can sample the survey itself to evaluate the
significance of galaxy condensations along the line-of-sight to lensing peaks.
The
sensitivity plot (Figure \ref{fig:sensitivity.ps}) 
suggests that we need not look deeper than
$z \sim 0.75$  to identify clusters corresponding to the three weak lensing peaks 
with $\nu > 4.25$ in the Subaru weak lensing map.
The most distant systems must be very rich and they should be obvious in the 
photometric data. We use the R$_C$-band counts in the image to check for such distant systems.

We examine the faint Subaru R$_C$ band counts (Miyazaki et al. 2007) in the range $21 < R_C < 23$ by binning the 
data in 0.3$^{\prime}$ bins and smoothing the map of counts with a 1.5$^{\prime}$ Gaussian
(the same smoothing and pixel scale as the Subaru weak lensing map).
The range in apparent magnitude just overlaps the $r_{petro} < 21.3$ magnitude limit of our main redshift survey sample.
The binning and smoothing lengths are essentially the same as those for the weak lensing map.
We define $\sigma_S$ as the standard deviation of the pixel value in the map of counts and we note 
peaks that rise $\gtrsim 2\sigma_S$ above the median as corroborating evidence for systems
of galaxies along the line-of-sight toward weak lensing peaks and/or as evidence for
systems beyond the reach of the redshift survey. Table \ref{tbl:VDisp} indicates the
lensing peak directions where there is a count excess. All three peaks with $\nu > 4.25$ correspond to count excesses.

Probes through the redshift survey centered on each galaxy in the survey
contain many of the
selection effects which impact a probe toward a lensing peak. Thus comparison
of these probes with similar probes toward the peaks should provide a 
measure of the significance of cluster candidates in the redshift
survey. 
We have demonstrated by experiment that  centering of probes on 
galaxies or on random points in the region has no effect on the results of
our analysis.

The steps in our candidate cluster identification procedure are:

\begin{enumerate}

\item We sample the redshift survey in  
{\it test cones} 
with  3$^{\prime}$ radius centered on each galaxy 
in the redshift survey. 
The cones are large enough to detect a galaxy cluster
across the redshift range we sample.

\item In each {\it test cone} we count 
the number of additional galaxies, N$_{gal}$, within
$0.004(1 + z)$ of the survey galaxy redshift. This window is comparable with the extent in redshift space of 
systems we wish to identify.

\item In each of the redshift bins of item 2, 
we evaluate the mean occupation and the
variance ($\sigma_{SH}$) across the entire survey. 
We then identify the set of {\it test cones} at each 
redshift which are 5$\sigma_{SH}$ above the mean occupation and contain at least
5 galaxies (N$_{gal} \geq 4$). The minimum number 
of galaxies enables computation of a
dispersion (albeit with large error). The 5$\sigma_{SH}$ limit restricts the
sample to high peaks which are reasonable candidate clusters especially at
the peak sensitivity of the weak lensing map, $z \sim 0.4$.
Both the 5 galaxy and 5$\sigma_{SH}$ limits are
generous; they admit many peaks well below the expected detection 
threshold for the weak lensing map (Figure \ref{fig:sensitivity.ps}).
We call these probes 
{\it 5$\sigma_{SH}$ probes} hereafter.

\item As a final consistency check we examine the Subaru images for evidence of
a massive system. We consider two indicators: (1) an $\gtrsim 2\sigma_S$ excess in the count of galaxies with
21$ < R_C < 23$, and (2) the presence of a visually identifiable red sequence.

\end {enumerate}

In Figure \ref{fig:sigmamap.scaled.ps} the 
two most significant weak lensing peaks (labeled 0 and 1) are coincident 
with two well populated clusters. These two clusters appear as prominent ``fingers'' at redshift 0.540 and 0.413, respectively,  in Figure 
\ref{fig:cone.diagram.subaru.ps}.

There are obviously many well populated 5$\sigma_{SH}$
probes without any associated significant weak lensing peak. 
Many of the well-populated redshift survey probes correspond to groups with small velocity dispersion.
From Figure \ref{fig:sensitivity.ps} we would not expect these systems with rest frame line-of-sight velocity dispersion $\sigma_{rf} \lesssim 500$ km s$^{-1}$ to produce a significant signal in the $\kappa$ map. There are also weak lensing peaks with no corresponding peak in the redshift survey.
It is interesting to note that none of the 5 unreliable peaks in Table \ref{tbl:VDisp} (starred entries)
coincide with well-populated probes in the redshift survey; removal of these unreliable peaks from consideration evidently increases the purity of the lensing map. Three of these unreliable peaks are above the $\nu = 4.25$ threshold.

To further assess the meaning of the weak lensing peaks we 
examine the redshift distributions along the line-of-sight 
toward significant weak lensing peaks. We ask which lines-of-sight intersect a
cluster with a velocity dispersion large enough to plausibly account for 
the weak lensing peak (Section \ref{clusters}).

\subsection {Convergence Map Peaks and Candidate Clusters}
\label{clusters}
In this section we evaluate candidate systems along the line-of-sight toward 
the 6 uncontaminated GTO2deg$^2$ weak lensing peaks with $\nu > 3.7$ 
in the revised Subaru weak lensing map (Section \ref{newmap}). The extension to the lower original significance limit of Miyazaki et al. (2007) is a consistency test of the implication of the B-mode that the threshold should be higher and that most of the peaks in the interval $3.7 < \nu < 4.25$ may be spurious (see e.g. Geller et al. 2010; Shan et al. 2011) This test can only be carried out with an independent  survey.
 
Occasionally a bright star can be accidentally superimposed near a massive cluster of galaxies or a cluster may be located near the edge of the survey field. We remove 5 weak lensing peaks that we regard as contaminated or, equivalently, unreliable because they are near a bright star or because they are too close to the survey boundary.
Neither
the redshift survey nor the faint galaxy counts reveal any candidate systems along the line-of-sight toward any of these peaks. Figure \ref{fig:velhisto3.ps} shows two examples of 
the redshift distributions toward these peaks; there are no candidate massive systems. The groups in the probe have rest frame line-of-sight velocity dispersions $\lesssim 250$ km s$^{-1}$.  Histograms for the other 3 contaminated lensing peaks are similar.  

The presence of high significance {\it a priori} unreliable peaks demonstrates that identifiable systematic effects can produce spurious high significance peaks in the weak lensing map.
Identification and removal of these unreliable contaminated peaks substantially increases the fidelity of the weak lensing map as a route to identifying massive systems of galaxies.

Figures \ref{fig:velhisto1.ps} and \ref{fig:velhisto2.ps} show the redshift $z$ distribution within 3$^\prime$ of the central position of each of the six uncontaminated candidate halos
with $\nu > 3.7$. Figure \ref{fig:velhisto1.ps} shows the three peaks with $\nu > 4.25$;
Figure \ref{fig:velhisto2.ps} shows the three peaks with $3.7 < \nu \leq 4.25$. The dark histogram shows the redshift distribution in 
bins of 0.002(1+$z$). The thin histogram shows the redshift distribution in a concentric cone with a $6^\prime$ radius in each of the candidate halo directions.

The probes toward peaks 0 and 1 in Figure \ref{fig:velhisto1.ps} each contain an impressive peak. These peaks correspond to obvious ``fingers'' in redshift space (Figure \ref{fig:velhisto1.ps}).  The rest frame line-of-sight velocity dispersions of these systems within a 6$^\prime$ probe are, respectively 576$\pm$64 km s$^{-1}$ and
598$\pm$58 km s$^{-1}$ in agreement with
previous measures (Dressler et al 1999 (peak 0); Hamana et al. 2008 (peak 1); see Section
\ref{history}).  Table \ref{tbl:VDisp} provides the number of galaxies and the rest frame line-of-sight velocity dispersion for both 3$^{\prime}$ and 6$^{\prime}$ probes. The rest frame line-of-sight velocity dispersions for these two clusters places both of them very near the
$\nu = 4.25$ detection threshold. Counts of faint galaxy within these probes show a
$> 2\sigma_S$  excess as one might expect (see Table \ref{tbl:VDisp}). 

Weak lensing detections of these clusters may be boosted by the obvious superposed structures in the redshift survey
(Hoekstra et al. 2001; White et al 2002; de Putter \& White 2005;  Hoekstra et al. 2011).
We note that the weak lensing map detects  the clusters corresponding to peaks 0 and 1 at
high significance $\nu \sim 6$. However for $\nu \sim 6$, our predicted velocity dispersions for these peaks exceed the measured values by only 20-30\%. The 1$\sigma$ errors in the velocity dispersions are $\gtrsim 10$\%. Thus we cannot draw any definitive conclusion for this apparent  difference between the measured velocity dispersion from the redshift survey and the weak lensing sensitivity analysis.

The probe toward weak lensing peak 3 shows a more complex situation. There are two significant peaks in the histogram, one at $z = 0.297$ and the other at $z = 0.673$. The line-of-sight velocity dispersions are  567 km s$^{-1}$ and 558 km s$^{-1}$, respectively within 3$^{\prime}$ probes; the rest frame line-of-sight velocity dispersions are systematically smaller in the 6$^{\prime}$ probes, but the errors are large. The faint galaxy counts show an excess (Table \ref{tbl:VDisp}), probably resulting from the more distant system. Both systems may contribute to  the lensing signal.  
We conclude that the purity for $\nu > 4.25$ is 2/3.

\begin{figure}[htb]
\centerline{\includegraphics[width=7.0in]{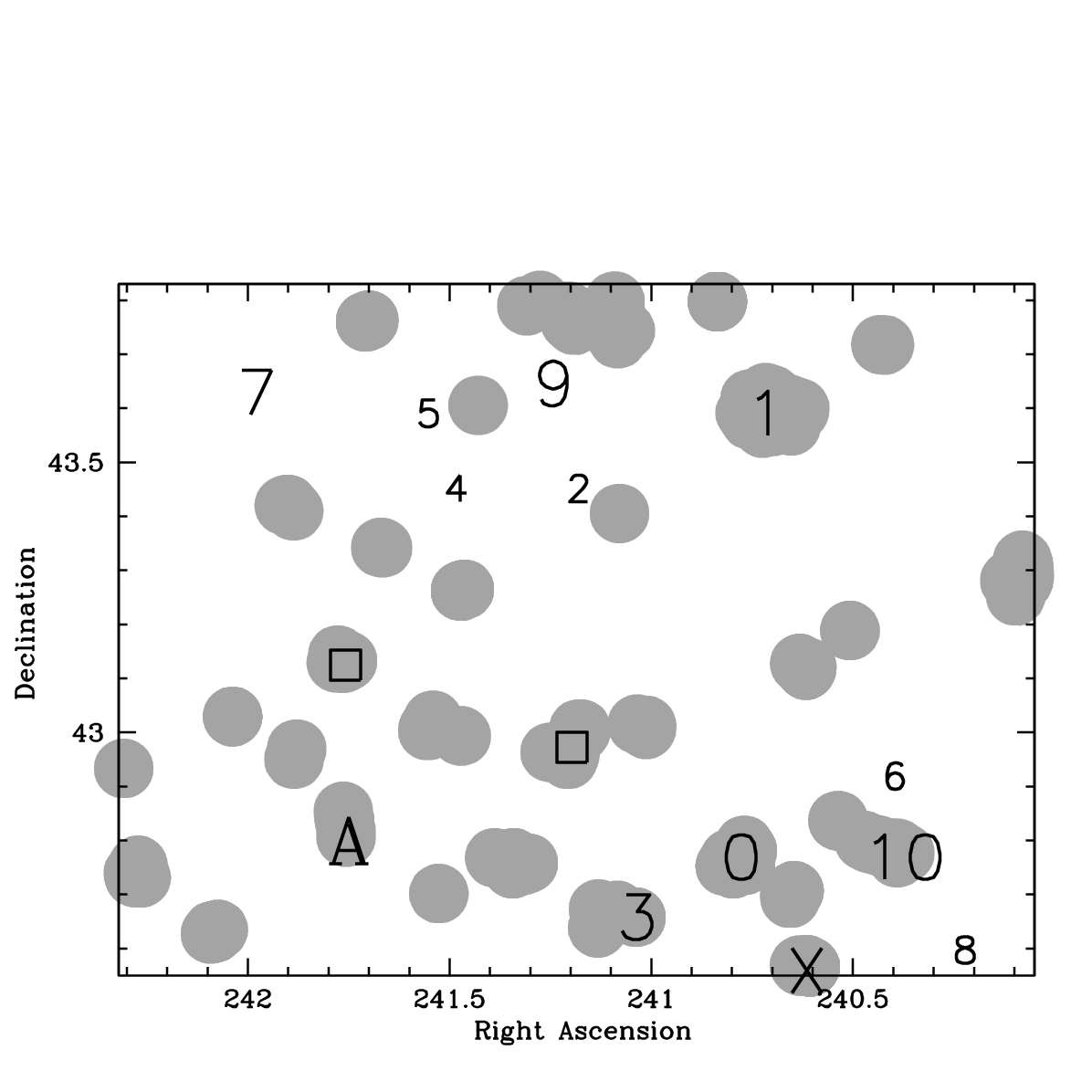}}
\vskip -5ex
\caption{Map of GTO2deg$^2$ centers of the 3$^{\prime}$ radius 5$\sigma_{SH}$ 
probes (Section \ref{sampling}; gray dots)  and centers of significant weak lensing peaks (large numbers are reliable peaks (no nearby bright stars or field edges); small numbers are unreliable). The value of the number gives the rank of the convergence map peak. The tick marks on the declination axis are 6$^\prime$ apart, the diameter of the probes. Note that reassuringly none of the unreliable peaks (2, 4, 5, 6, and 8) overlap the gray dots. The X marks the position of a cluster detected in the redshift survey, but it is too close to the edge of the survey for reliable detection in the lensing map; open squares are sample directions which are well-populated in the redshift survey but where there is no weak lensing peak significant at $\nu \gtrsim 2$; A denotes
the position of the Abell cluster 2158.
\label{fig:sigmamap.scaled.ps}}
\end{figure}

\begin{figure}[htb]
\centerline{\includegraphics[width=7.0in]{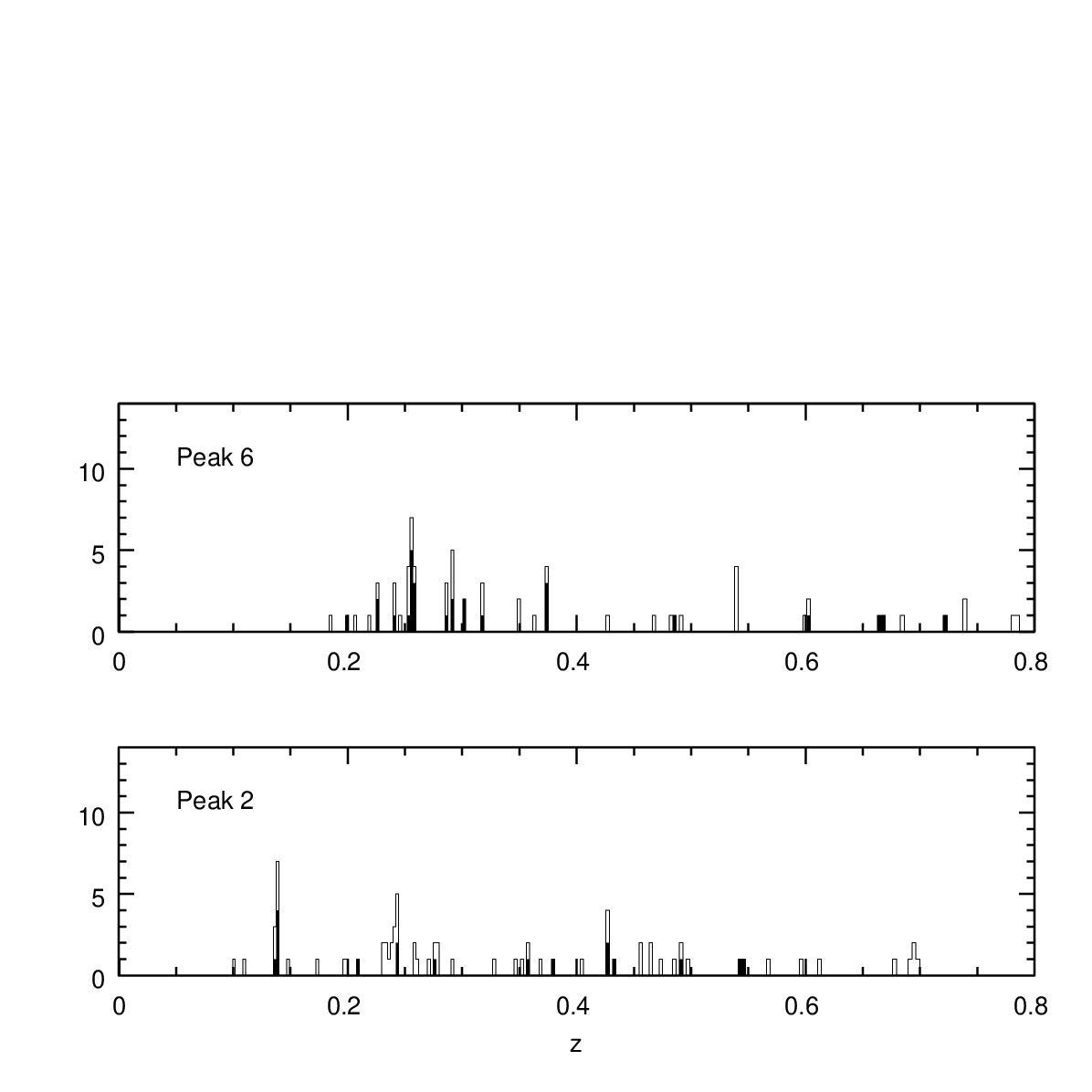}}
\vskip -5ex
\caption{Redshift histograms for cones toward two weak lensing peaks with nearby bright stars or field edges in Table \ref{tbl:VDisp}. Peak 2 has $\nu > 4.25$; peak 6 has $3.7 < \nu \leq 4.25$
The solid histogram shows the redshift distribution for a cone with a 3$^\prime$ radius; the open histogram shows the distribution for a 6$^{\prime}$ radius. Bins in redshift are 
$0.002(1 + z)$ wide. These histograms confirm the impression from Figure \ref{fig:cone.photoZ.ps} that there are no systems of galaxies along the lines-of-sight toward these peaks resulting from obvious systematic errors in the map.
\label{fig:velhisto3.ps}}
\end{figure}

\begin{figure}[htb]
\centerline{\includegraphics[width=7.0in]{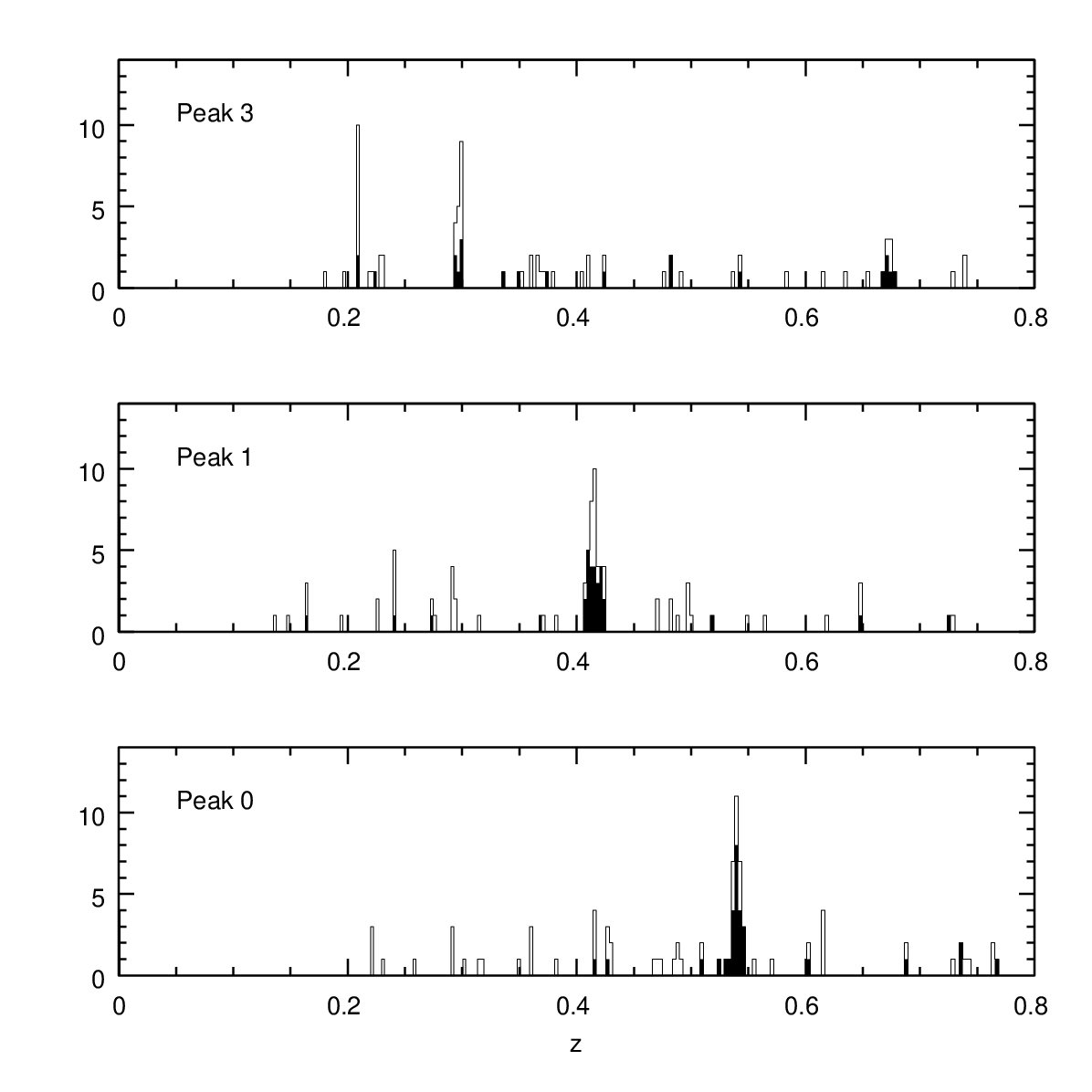}}
\vskip -5ex
\caption{Redshift histograms for cones toward the reliable (without nearby bright stars or field edges) weak lensing peaks with
$\nu > 4.25$ in Table \ref{tbl:VDisp}.
The solid histogram shows the redshift distribution for a cone with a 3$^\prime$ radius; the open histogram shows the distribution for a 6$^{\prime}$ radius. Bins in redshift are $0.002(1 + z)$ wide. 
\label{fig:velhisto1.ps}}
\end{figure}

\begin{figure}[htb]
\centerline{\includegraphics[width=7.0in]{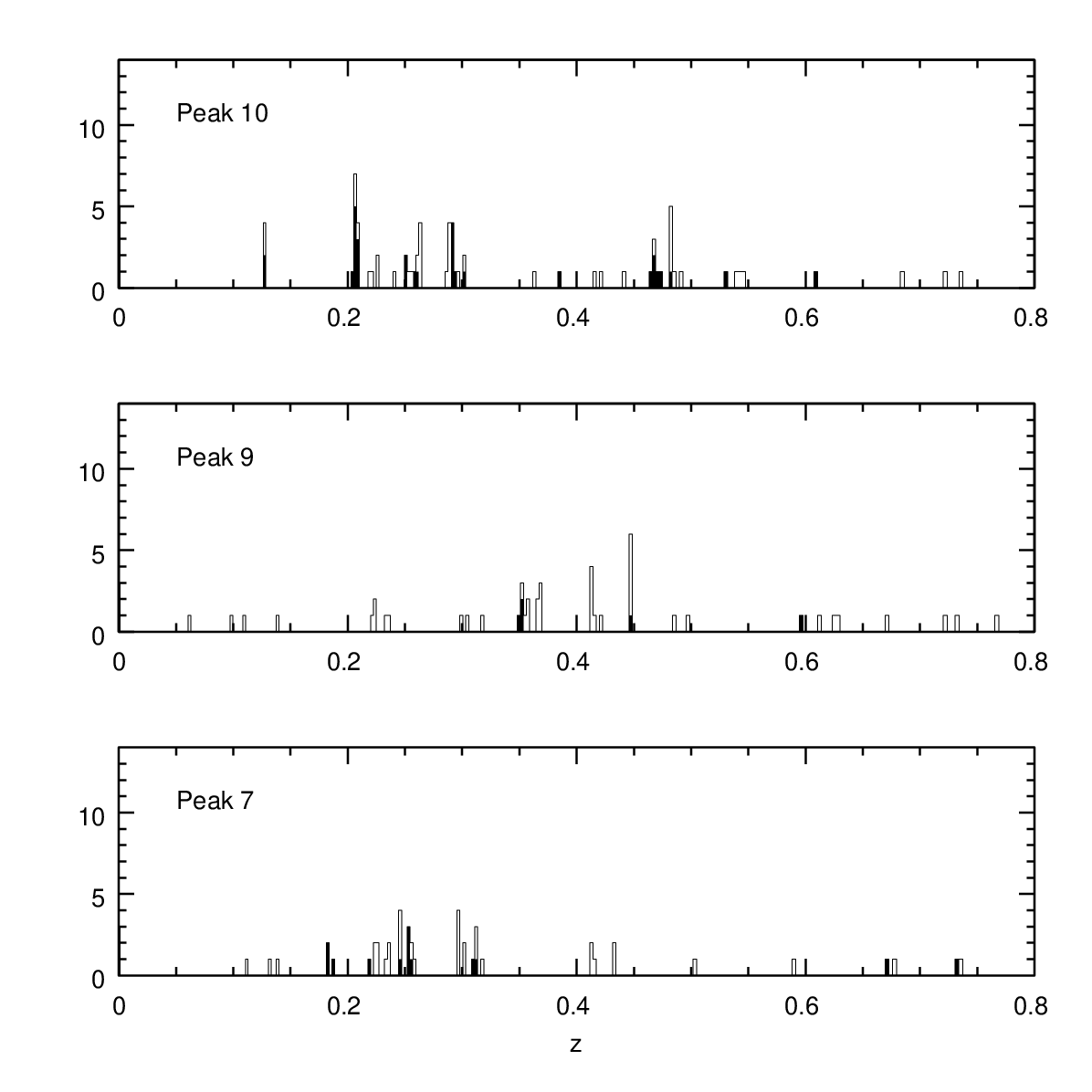}}
\vskip -5ex
\caption{Redshift histograms for cones toward  weak lensing peaks in Table \ref{tbl:VDisp} with $3.7 < \nu \leq 4.25$; these peaks are not contaminated by neaby bright stars or field edges.
The solid histogram shows the redshift distribution for a cone with a 3$^\prime$ radius; the open histogram shows the distribution for a 6$^{\prime}$ radius. Bins in redshift are 
$0.002(1 + z)$ wide.
\label{fig:velhisto2.ps}}
\end{figure}

In Figure \ref{fig:velhisto2.ps}, probes toward weak lensing peaks 7 and 9 reveal no candidate systems. Figure \ref{fig:sigmamap.scaled.ps} shows that neither of these peaks overlaps a well-populated probe through the redshift survey. Furthermore, there
is no excess in the faint galaxy counts. 

Along the line-of-sight toward weak lensing peak 10 (Figure \ref{fig:velhisto2.ps}), there are several narrow peaks corresponding to superposed groups along the line-of-sight.  
There is  a group at $z = 0.469$ with a velocity dispersion of 545 km
s$^{-1}$ within a 6$^{\prime}$ probe. We identify only 6 members of this system. The velocity dispersion of this system places it near the detection threshold albeit with a very large error. The most populated system along this line-of-sight  (as a result of the selection function of the survey) has a mean redshift 
$z = 0.208$ and a line-of-sight velocity dispersion of 404 km s$^{-1}$. This system corresponds to a MaxBCG cluster (Koester et al. 2007)
at $\alpha_{2000} = 240.39$, $\delta_{2000}= 42.76$ with a quoted photoz of 0.254
and a BCG spectroscopic redshift of 0.208. This system may enhance the lensing signal from the more distant system. There is a faint count excess in this direction Table (\ref{tbl:VDisp}). 

In the B-mode map there are 3 uncontaminated peaks with $3.7 < \nu \leq 4.25$; in the $\kappa$-map there are also 3 uncontaminated peaks. At most one of the $\kappa$-S/N map peaks in this significance range corresponds to a rich system of galaxies or, more probably, a superposition along the line-of-sight, consistent with expectations based on the B-mode map. Geller et al (2010) and Shan et al (2011) also find that the purity of weak lensing candidate halo samples declines with decreasing significance probably consistent with expectations based on the B-mode maps.

\begin{figure}[htb]
\vskip -0.5in
\centerline{\includegraphics[width=7.0in]{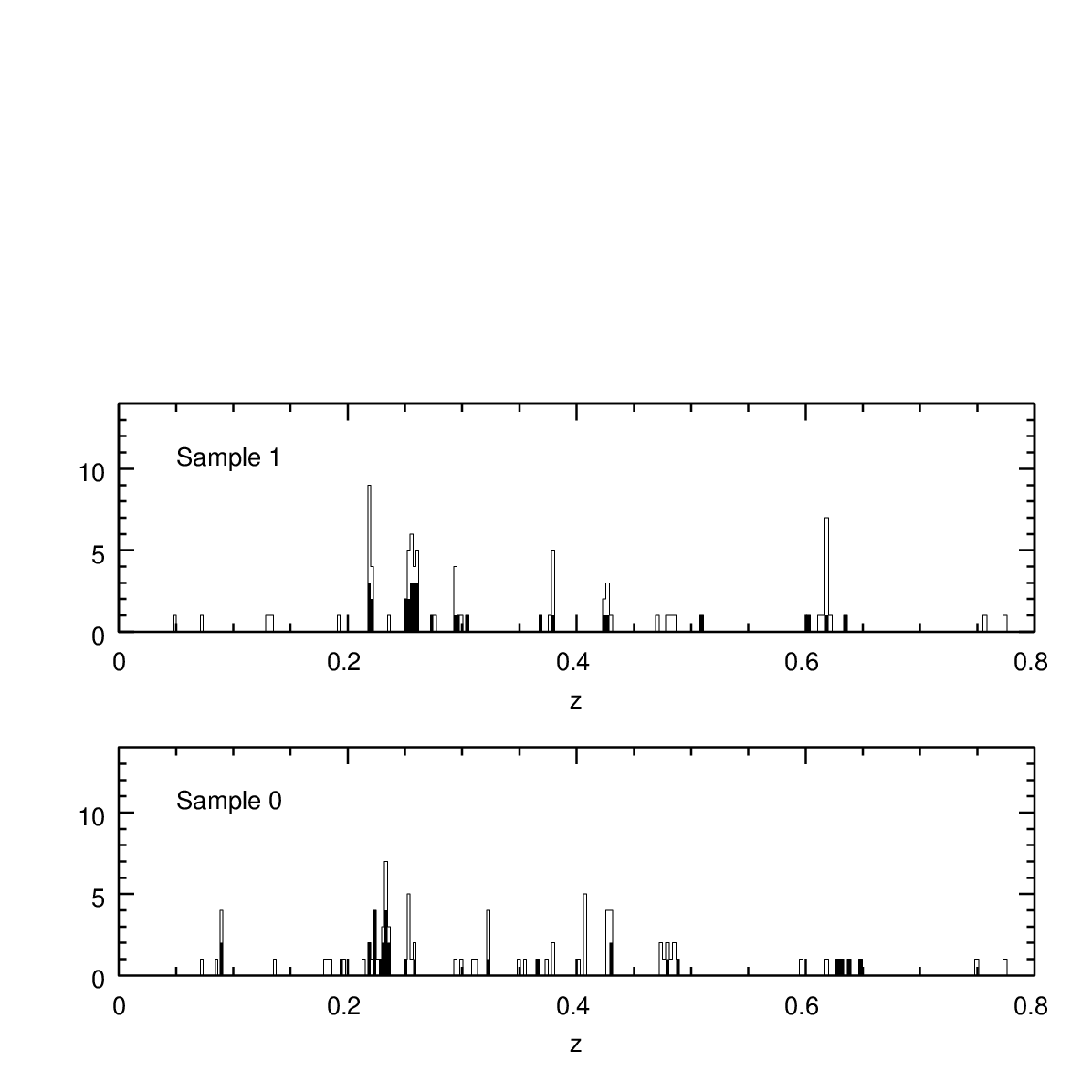}}
\vskip -5ex
\caption{Redshift histograms for cones in two directions
below the 2$\sigma_{\kappa-S/N}$ contour of the weak lensing map. 
The solid histogram shows the redshift distribution for a cone with a 3$^\prime$ radius; the open histogram shows the distribution for a 6$^{\prime}$ radius. Bins in redshift are 
$0.002(1 + z)$ wide. Groups with rest frame line-of-sight velocity dispersions $\lesssim 400$ km s$^{-1}$ populate these probes. They are actually more densely populated than probes toward the spurious  weak lensing peaks 7 and 9 (Figure \ref{fig:velhisto3.ps}).
\label{fig:velhisto4.ps}}
\end{figure}

As another test of the nature of the  weak lensing map, we examine probes through the redshift survey toward the two well-populated positions indicated by heavy squares in Figures \ref{fig:GTOkappawopeak.ps} and \ref{fig:sigmamap.scaled.ps}. In the weak lensing map, there are no peaks significant at $\nu \gtrsim 2$ that are coincident with these redshift survey positions. Figure \ref {fig:velhisto4.ps} shows probes through the redshift survey in these two directions. By selection, the probes contain more galaxies than the probes toward
the spurious weak lensing peaks 7 and 9 with $3.7 < \nu \leq 4.25$. They are populated by a superposition of groups with line-of-sight velocity dispersions $\lesssim 400$ km s$^{-1}$. 

In summary, the 3 weak lensing candidate halos with $\nu > 4.25$ yield a sample of 2 rich clusters associated with peaks 0 and 1 and a superposition associated with peak 3. Thus even at this high threshold, the purity is 2/3. for comparison, using the same approach, the purity of the DLS is 3/3 for $\nu > 4.25$ (Table \ref{tbl:Maps}; Geller et al. (2010)). Of course, the uncertainty in these consistent results is large because of the small samples of peaks.

\subsection{Completeness and Additional GTO2deg$^2$ Redshift Survey Candidate Clusters}
\label{zclusters}

The redshift survey provides a route to identifying candidate galaxy systems undetected by the weak lensing map. It thus provides some assessment of the completeness of the cluster candidate list derived from the weak lensing map. There are many approaches to 
identifying candidate systems in the redshift survey. For consistency we use redshift survey probes to identify systems in the same way that we use them to test for systems along the line-of-sight toward weak lensing peaks. We restrict the analysis to systems that should be detected at or above our 4.25$\sigma_{\kappa - S/N}$ threshold.

Figure \ref{fig:sigmamap.scaled.ps} shows the positions of all of the 5$\sigma_{SH}$ probes through the redshift survey; these probes contain a well-populated redshift bin
(see Section \ref{sampling}). Of course the bins are large enough that 
they may contain a superposition of poor groups rather than a richer system. We compute
line-of-sight velocity dispersions for the peaks within each probe. In addition to the
clusters corresponding to weak lensing peaks 0 and 1, we find only one additional  
concentration in redshift space that corresponds to a massive system. 

\begin{figure}[htb]

\centerline{\includegraphics[width=3.5in, angle = -90]{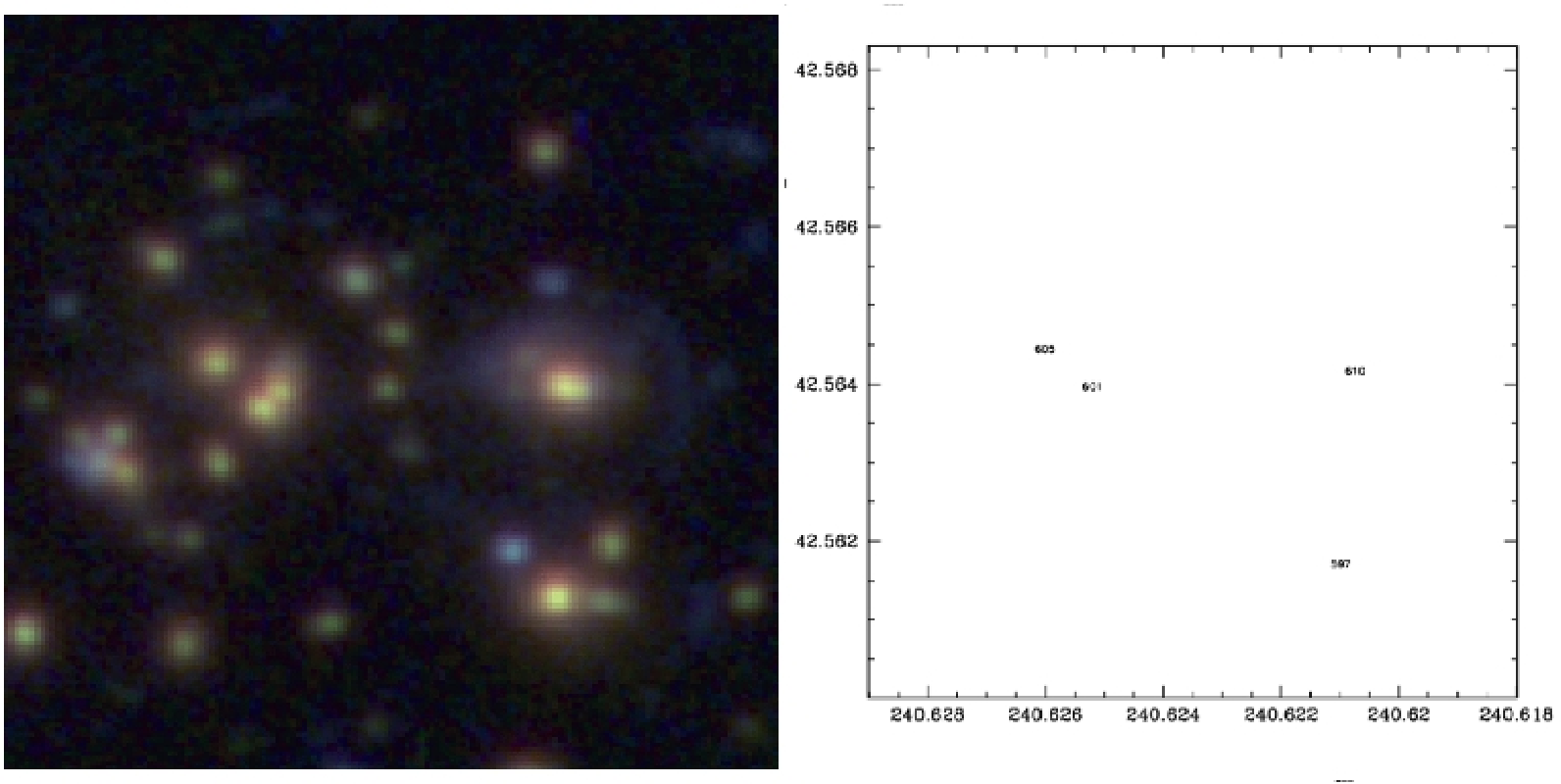}}
\vskip 7ex
\caption{Densest region of the cluster discovered in the redshift survey at $z = 0.602$. Note the faint blue arc to the right of the brightest cluster member
(labeled 610 in the right-hand panel). The left panel shows the Subaru image of the cluster center; the right panel shows the
positions of galaxies with redshifts (numbers indicate 1000*z for the galaxy).
\label{fig:arc.ps}}
\end{figure}

The cluster we identify from the redshift survey at $z = 0.602$ has an impressively dense central region and the Subaru image reveals at least one faint strong lensing arc
(Figure \ref{fig:arc.ps}). Detection in the weak lensing map is compromised 
because the cluster center ($\alpha_{2000}$ = 240.62; $\delta_{2000}$ = 42.57) is only $\sim 1.2$ arcminutes from the edge of the field. The
weak lensing map does have a peak coincident with this cluster at a significance $\nu = 2.9$ (We list this cluster along with an estimate of its line-of-sight velocity dispersion as the  last entry in Table \ref{tbl:VDisp}).

Examination of the well-populated redshift survey probes also reveals a dense system with
a rest frame line-of-sight velocity dispersion of 419$\pm$24 km s$^{-1}$ and a mean redshift of 0.185 based on 23 members within 6$^\prime$ of the cluster center ($\alpha_{2000}$ = 241.76, $\delta_{2000}$ = 42.80). 
This system is obvious as a short, well-defined ``finger'' in Figure \ref {fig:cone.diagram.subaru.ps}. We identify it with the richness class zero Abell cluster 2158 (A2158C: Struble \& Rood 1999; see also Abell 1958). Foreground groups at a redshift $ z = 0.13$ have confused the identification of this system in the past. In the lensing map this cluster appears
below our $\nu = 4.25$  detection  threshold;  it is coincident with a weak lensing peak significant at $\nu = 3.4 $. (We list this cluster as the next to last entry in Table \ref{tbl:VDisp}).

Other than the clusters corresponding to weak lensing peaks 0 and 1, we identify no clusters in the redshift survey that should produce a weak lensing signal $\gtrsim 4.25 \sigma_{\kappa -S/N}$. Thus, the set of cluster detections appears to be complete.

\section {Comparing the Weak Lensing Sensitivity of the GTO2deg$^2$ and DLS F2 Fields}
\label{comparison}

The effects of atmospheric seeing are a challenge in constructing weak lensing maps (Schneider 2006). Because the source galaxies are small compared with typical atmospheric seeing, the signal-to-noise of a weak lensing map is sensitive to the seeing.
Schneider emphasizes the importance of good seeing  for the determination of source ellipticities once a PSF correction is made. Schneider (2006) argues that at fixed redshift, a shallower map with better seeing should enable weak lensing detection of lower mass (lower velocity dispersion) systems compared with a deeper map in less good seeing. We examine that conjecture next.  

\begin{figure}[htb]
\centerline{\includegraphics[width=6.7in]{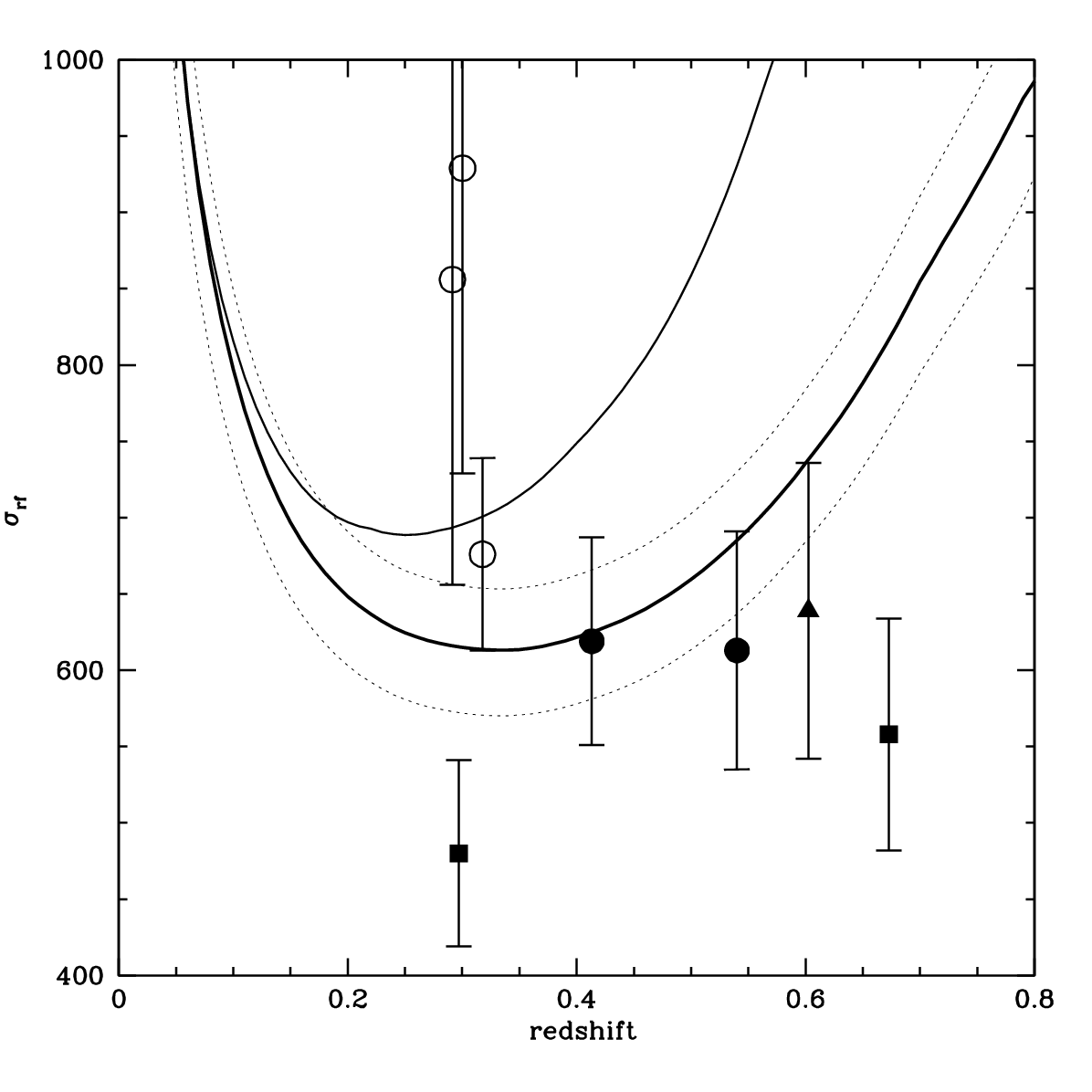}}
\vskip -5ex
\caption{Comparison of cluster detection and sensitivity of the GTO2deg$^2$ and DLS F2 weak lensing maps. The curves show the $\nu = 4.25$ detection threshold for the two maps.
The heavy curve refers to the GTO2deg$^2$ field; the lighter curve refers to the DLS F2 field. Open points with error bars are the three clusters detected in the DLS lensing map and redshift survey. The solid points are groups and clusters in the directions of the GTO2deg$^2$ weak lensing peaks: solid circles are peaks 0 and 1; solid squares are the superposition of peak 3. The solid triangle indicates the cluster at redshift 0.602 discovered in the redshift survey but too near the edge of  the Subaru field for reliable weak lensing detection.
\label{fig:sensitive2.ps}}
\end{figure}

The GTO2deg$^2$ field consists of subfields with a 30-40 minute exposure on the Subaru 8.2-meter telescope in  0.66$^{\prime\prime}$ seeing; the DLS F2 field is a 5-hour exposure on the KPNO 4-meter in 0.85$^{\prime\prime}$ seeing. Figure \ref{fig:sensitive2.ps} compares the weak lensing sensitivity of the two fields. The curves show the rest frame line-of sight velocity dispersion for a system which could be detected at  a $\nu = 4.25$ at each redshift. At every redshift, the
Subaru observations have the potential to yield a  detection at a lower line-of-sight velocity dispersion. The maximum sensitivity is shifted toward greater redshift for the Subaru data. Table \ref{tbl:Maps} shows the substantial difference in the number of resolved sources for the two weak lensing maps: 22.4  arcminute$^{-2}$ for the DLS and 35 arcminute$^{-2}$ for the GTO2deg$^2$ field. Table \ref{tbl:Maps} lists other relevant parameters for the two surveys. In constructing Table \ref{tbl:Maps} we have reevaluated the
DLS F2 results of Geller et al. (2010) using a $\nu = 4.25$ rather than a $\nu =3.5$ threshold for selecting significant peaks in the weak lensing map. We have also included the
Miyazaki et al. (2007) procedure for removing unreliable, but apparently high significance weak lensing peaks. 

Figure \ref{fig:sensitive2.ps} also shows the rest frame line-of-sight velocity dispersions for single massive systems in the
foreground redshift surveys corresponding to weak lensing peaks with $\nu > $ 4.25 in both surveys. The most telling and suggestive difference between the admittedly small samples of clusters in the two surveys is the demonstration that, as expected
from the sensitivity curves, the Subaru map cleanly detects clusters with rest frame line-of sight velocity significantly below the limits of the DLS sensitivity at a fixed redshift.

Understanding both the completeness and the purity of weak lensing map cluster detections are, of course, crucial to their use as cosmological tools for measuring e.g. the cluster mass function (White et al. 2002; Hennawi \& Spergel 2005). The purity of weak lensing cluster detection is consistent with the more pessimistic theoretical models (White et al. 2002; Hamana et al. 2004; Hennawi \& Spergel 2005;
Maturi et al. 2010). even for $\nu > 4.25$, The DLS weak lensing cluster sample is $\sim 50$\% complete relative to a sample drawn from the foreground SHELS redshift survey (Geller et al. 2010).
The GTO2deg$^2$ cluster catalog is complete (Section \ref{clusters}), but the sample is so small that we cannot draw a general conclusion. These samples certainly emphasize the
need for more extensive detailed tests not dominated by small number statistics. The CFHTLS
(Shan et al. (2011) makes substantial progress toward this goal, but the analysis is based on photometric redshifts rather than on a redshift survey that can separate superpositions along the line-of-sight unresolved by photometric redshifts.

\section {Conclusion}
\label{conclusion}

Weak lensing surveys are a potential tool for identifying massive systems of galaxies. These mass selected cluster catalogs could provide important cosmological constraints. Objective tests of the completeness and purity are an important foundation the application of future surveys including DES, LSST, and EUCLID to the construction of catalogs of cluster candidates from weak lensing maps. 

The Subaru GTO2deg$^2$ field (centered at $\alpha_{2000}$ = 16$^h$04$^m$44$^s$;$\delta_{2000}$ =43$^\circ$11$^{\prime}$24$^{\prime\prime}$) is the second where we have carried out a dense 
foreground redshift survey as a test of the fidelity of the weak lensing map. In contrast with the 0.85$^{\prime\prime}$ seeing of the DLS F2 field of our first survey (Geller et al. 2010), the Subaru field has median seeing of 0.66$^{\prime\prime}$. The number of resolved sources in the Subaru field is 35 arcmin$^{-2}$ in contrast with 22 arcmin$^{-2}$ for the DLS field. 

To match the sensitivity of the Subaru lensing map, the primary component of our redshift survey includes red galaxies ($g -r > 1.0$ and $r-i > 0.4$) with SDSS Petrosian $17.77 < r < 21.3$. The total number of galaxies with redshifts in the field is 4959: 4541 are new
redshifts measured with Hectospec. This dense survey enables a variety of tests of the weak lensing map.

Our tests include (1) searching the redshift survey for massive systems along the 
line-of-sight toward weak lensing peaks with significance $\nu > 3.7$, the threshold originally used by Miyazaki et al (2007), (2) searching the redshift survey for all systems massive enough to produce a lensing signal to estimate the completeness of the weak lensing detections and (3) examining well-populated regions of the redshift survey in directions where there is no significant weak lensing signal to
further test the fidelity of the weak lensing map.

We show that the procedure Miyazaki et al. (2007) used to eliminate contaminated, but apparently high confidence peaks from their weak lensing map improves the fidelity of the weak lensing map in comparison with the foreground redshift survey. These contaminated peaks should probably be removed from both E- and B-mode maps prior to analysis.

The B-mode map derived from a reanalysis of the Subaru data suggest that $\nu =4.25$ is a better cluster detection threshold than $\nu = 3.7$. Comparison with the foreground redshift survey confirms the implications of the B-mode map.
In the Subaru map,  2/3 of the weak lensing peaks with $\nu > 4.25$ are detections of individual massive systems. The third peak is a superpositon of of less massive systems. In the larger DLS F2 field, a deeper observation in poorer seeing (Table \ref{tbl:Maps}) all three of the $\nu > 4.25$ peaks correspond to massive systems of galaxies. Thus
for $\nu > 4.25$, the purity is $\gtrsim 67$\%.

In the DLS F2 field, the SHELS foreground redshift survey identifies twice as many massive cluster candidates as the weak lensing map; in the Subaru field all of the cluster candidates are detected in both the redshift survey and the weak lensing map, but the sample is very small. These small samples
suggest that for $\nu > 4.25$ the completeness is $\gtrsim 50$\%.

In contrast with the DLS F2 field, the Subaru map detects less massive systems at a fixed redshift as one would expect from the greater source density. This comparison suggests that
robust measurements of the cluster mass function down to a mass of $1.7 \times 10^{14}$ M$_\odot$  for redshifts $\lesssim 0.6$ can be derived from weak lensing maps provided the purity and completeness 
could be understood.

Taken together, the Subaru GTO2deg$^2$ and DLS F2 fields underscore the need for a large-area direct test of weak lensing cluster identification based on high significance peaks
derived from weak lensing.  The CFHTLS (Shan et al. (2011) is an important step in this direction.
Well-designed foreground redshift surveys of this region will refine the correspondence (or lack of it) between systems of galaxies and weak lensing peaks.

\acknowledgments
We thank P. Berlind and M. Calkins for their expert operation
of the Hectospec.  D. Mink, J. Roll, S. Tokarz, and W. Wyatt
constructed and ran the Hectospec pipeline. Nelson Caldwell deftly
manages Hectospec queue scheduling for optimal scientific results. Eduard Westra assisted with
some of the observations. We thank Scott Kenyon for insightful comments and discussions. We thank the anonymous referee for astute comments that improved this paper.
The Smithsonian
Institution generously supported Hectospec and this project. NSF grant 
AST-0708433 supports Ian Dell Antonio's research.

{\it Facilities:}\facility {MMT(Hectospec)}

\clearpage
\begin{landscape}
\begin{deluxetable}{rrrrrrccc} 
\tablecolumns{9} 
\tablewidth{0pc} 
\tabletypesize{\footnotesize}
\tablenum{1}
\tablecaption{Redshifts in the GTO2deg$^2$ Field} 
\tablehead{ 
\colhead{objID} & \colhead{ra}   & \colhead{dec}    & \colhead{redshift} & 
\colhead{z error}    & \colhead{$r_{petro}$}   & \colhead{$(g-r)_{fiber}$}    & \colhead{$(r-i)_{fiber}$} & \colhead{zSource}}
\startdata 
587733442122744127 & 240.0929 & 42.7913 & 0.12636 & 0.00009 &  18.97 &   0.94 &   0.52 & Hecto \\
587733442122744638 & 240.0663 & 42.8125 & 0.48417 & 0.00011 &  20.86 &   1.02 &   0.49 & Hecto \\
587733442122745038 & 240.0662 & 42.7705 & 0.68305 & 0.00020 &  22.31 &   1.35 &   1.15 & Hecto \\
587733442122745075 & 240.0767 & 42.7428 & 0.68472 & 0.00022 &  21.03 &   1.55 &   1.20 & Hecto \\
587733442122809425 & 240.1192 & 42.6976 & 0.59909 & 0.00021 &  20.80 &   1.22 &   0.89 & Hecto \\
587733442122809463 & 240.2103 & 42.7137 & 0.48322 & 0.00013 &  20.14 &   1.32 &   0.81 & Hecto \\
587733442122809491 & 240.1748 & 42.6516 & 0.46796 & 0.00012 &  20.46 &   1.87 &   0.87 & Hecto \\
587733442122809551 & 240.0543 & 42.6959 & 0.09460 & 0.00006 &  17.64 &   0.89 &   0.41 & Hecto \\

\enddata 
\label{tbl:redshifts}
\end{deluxetable}
 
\end{landscape}

\clearpage
\begin{deluxetable}{lcc}
\tablecolumns{3}
\tablewidth{0pc}
\tabletypesize{\footnotesize}
\tablenum{2}
\tablecaption{Comparison of GTO2deg$^2$ and DLS F2 Field Weak Lensing Maps}
\tablehead{
\colhead{Property}&
\colhead{GTO2deg$^2$}&
\colhead{DLS F2}
}
\startdata 
Field Center ($\alpha_{2000}, \delta_{2000}$)&241.183, 43.190&139.885, 30.000 \\
Field Area(deg$^2$)&2.02 & 4.1\\
Source Density (arcmin$^{-2}$)& 35&22.4\\
Median Seeing ($^{\prime\prime}$)&0.66 &0.85\\
Seeing Range ($^{\prime\prime}$)&0.63-0.76  &$0.82-0.90$\\
Pixel Scale ($^{\prime\prime}$)&18&30\\
Smoothing Length (arcmin)&1.5 &1.5\\
N$_{peaks} > 4.25$&3 &3\\
Clusters detected&2&3\\
\enddata
\label{tbl:Maps}
\end{deluxetable}

\clearpage
\begin{landscape}
\begin{deluxetable}{lccccccccccc}
\tablecolumns{12}
\tablewidth{0pc}
\tabletypesize{\scriptsize}
\tablenum{3}
\tablecaption {Velocity Dispersions of GTO 2deg$^2$ Weak Lensing Candidate Clusters\tablenotemark{1}}
\tablehead{
\colhead {Rank}&
\colhead {$\nu$}&
\colhead {RA$_{2000}$}&
\colhead {DEC$_{2000}$}&
\colhead {z}&
\colhead {$\sigma_{rf,3}$ }&
\colhead {N$_3$}&
\colhead {err$_3$}&
\colhead {$\sigma_{rf,6}$}&
\colhead {N$_6$}&
\colhead {err$_6$} &
\colhead {Count}\\  & & & & & 
\colhead {km/s}& &
\colhead {km/s}&
\colhead {km/s}& &
\colhead {km/s} &
\colhead {Excess$^4$}

}
\startdata 
0&5.784520&240.776778&42.769138&0.540 & 613&20 &78 &576&29 &64&y \\
1 &5.673880& 240.715968&43.590494&0.413&619 & 15&68 &598&26&58&y  \\
2$^*$&4.451760&241.179641&43.451752& & & & & & & &\\
3& 4.426710&241.035219&42.658144&0.297 &595 &6& 87&480&18 & 61&y\\
& & & &0.673&558&5&76&431&8&77& \\
4$^*$& 4.383750&241.483689&43.451752& & & & & & & &\\
5$^*$&4.352070&241.552100&43.590494& & & & & & & &\\
6$^*$& 4.201050&240.396718&42.918980& & & & & & & &\\
7&3.819900&241.977766&43.630180& & & & & & & &\\
8$^*$&3.802810&240.221891&42.597097& & & & & & & &\\
9&3.784870&241.240451&43.645991& & & & & & & &\\
10&3.734410&240.373915&42.769138&0.469&595&5 &180 &545 &6 & 156& y\\
14$^2$&3.4&241.757&42.791&0.185&463&16&58&419&24&48&\\
37$^3$&2.9&240.616&42.569&0.602&639&8&97&639&8&97&\\
\enddata

%
\tablenotetext{*}{These weak lensing map peaks are not secure because they are within
4 arcminutes of a bright star or within 2.7 arcminutes of the survey boundary.}
\tablenotetext{1}{The Table headings $\sigma_{rf,3}$, N$_3$, and err$_3$ refer to the rest-frame line-of-sight velocity dispersion, the number of galaxies, and the error in the velocity dispersion for the 3 arcmin samples. Analogous quantities with subscript `6' refer to the 6 arcmin samples.}
\tablenotetext{2}{This cluster is Abell 2158.}
\tablenotetext{3}{We identify this system from the redshift survey alone, but it is coincident with a weak lensing peak significant at $\nu = 2.9 $. The lensing peak is not secure because the cluster is too close to the survey boundary.}
\tablenotetext{4}{A ``y'' means there is a $>2\sigma_S$ excess in the faint galaxy count
for $21 < R_C < 23$ in the direction of the weak lensing peak.}
\label{tbl:VDisp}
\end{deluxetable}
\end{landscape}

\clearpage
\begin{deluxetable}{lccc}
\tablecolumns{4}
\tablewidth{0pc}
\tabletypesize{\footnotesize}
\tablenum{4}
\tablecaption{B-Mode Map Peaks}
\tablehead{
\colhead{Peak ID}&
\colhead{RA$_{2000}$}&
\colhead{Dec$_{2000}$}&
\colhead{S/N}
}
\startdata
B0 & 240.9592 & 43.7292 &     4.2373 \\
B1$^*$ & 241.2405 & 43.5516 &     3.8413 \\
B2 & 240.9896 & 43.6682 &     3.8035 \\
B3 & 241.5141 & 42.9023 &     3.7122 \\
\enddata

\tablenotetext{*}{This B-mode  map peaks is not secure (in the same sense as the $^*$ peaks in the $\kappa$-S/N map) because it is within 4 arcminutes of a bright star.}
\label{Bmodetab}
\end{deluxetable}

\clearpage
\clearpage

\clearpage


\clearpage
\begin{thebibliography}{99}


\bibitem[Abell(1958)]{Abell58} Abell, G.~O.\ 1958, \apjs, 3, 
211 

\bibitem[Adelman-McCarthy et al.(2008)]{SDSS} 
Adelman-McCarthy, J.~K., et al.\ 2008, \apjs, 175, 297 

\bibitem[Bellagamba et al.(2011)]{2011MNRAS.413.1145B} Bellagamba, F., 
Maturi, M., Hamana, T., et al.\ 2011, \mnras, 413, 1145 




\bibitem[Berg{\'e} et al.(2008)]{Berge08} Berg{\'e}, J., et 
al.\ 2008, \mnras, 385, 695 


\bibitem[de Putter 
\& White(2005)]{dePutter05} de Putter, R., \& White, M.\ 2005, \na, 10, 676 



\bibitem[Dietrich et al.(2007)]{dietrich07} Dietrich, J.~P., Erben, T., Lamer, G., Schneider, P., Schwope, A., Hartlap, J., \& Maturi, M.\ 2007, \aap, 470, 821 

\bibitem[Casertano et al.(2000)]{casertano00} Casertano, S., et 
al.\ 2000, \aj, 120, 2747 

\bibitem[Dressler et al.(1999)]{dressler99} Dressler, A., Smail, 
I., Poggianti, B.~M., Butcher, H., Couch, W.~J., Ellis, R.~S., 
\& Oemler, A., Jr.\ 1999, \apjs, 122, 51 
 


\bibitem[Fabricant et al.(1998)]{Fabricant98} Fabricant, D.~G., 
Hertz, E.~N., Szentgyorgyi, A.~H., Fata, R.~G., Roll, J.~B., 
\& Zajac, J.~M.\ 1998, \procspie, 3355, 285 

\bibitem[Fabricant et al.(2005)]{Fabricant05} Fabricant, D., et 
al.\ 2005, \pasp, 117, 1411 

\bibitem[Gavazzi 
\& Soucail(2007)]{Gavazzi07} Gavazzi, R., \& Soucail, G.
\ 2007, \aap, 462, 459 

\bibitem[Geller et al.(2005)]{Geller05} Geller, M.~J., 
Dell'Antonio, I.~P., Kurtz, M.~J., Ramella, M., Fabricant, D.~G., Caldwell, 
N., Tyson, J.~A., \& Wittman, D.\ 2005, \apjl, 635, L125 

\bibitem[Geller et al.(2010)]{2010ApJ...709..832G} Geller, M.~J., Kurtz, 
M.~J., Dell'Antonio, I.~P., Ramella, M., 
\& Fabricant, D.~G.\ 2010, \apj, 709, 832 



\bibitem[Girardi 
\& Mezzetti(2001)]{girardi01} Girardi, M., \& Mezzetti, M.\ 2001, \apj, 548, 79 

\bibitem[Gunn et al.(1986)]{1986ApJ...306...30G} Gunn, J.~E., Hoessel, 
J.~G., \& Oke, J.~B.\ 1986, \apj, 306, 30 




\bibitem[Hamana et al.(2004)]{Hamana04} Hamana, T., Takada, M., 
\& Yoshida, N.\ 2004, \mnras, 350, 893 


\bibitem[Hamana et al.(2009)]{hamana09} Hamana, T., Miyazaki, 
S., Kashikawa, N., Ellis, R.~S., Massey, R.~J., Refregier, A., 
\& Taylor, J.~E.\ 2009, \pasj, 61, 833 

\bibitem[Hennawi 
\& Spergel(2005)]{hennawi05} Hennawi, J.~F., \& Spergel, D.~N.\ 2005, \apj, 624, 59 

\bibitem[Hoekstra(2001)]{Hoekstra01} Hoekstra, H.\ 2001, \aap, 370, 743 


\bibitem[Hoekstra et al.(1998)]{1998ApJ...504..636H} Hoekstra, H., Franx, 
M., Kuijken, K., \& Squires, G.\ 1998, \apj, 504, 636 

\bibitem[Hoekstra et al.(2011)]{Hoekstra11} Hoekstra, H., Hartlap, 
J., Hilbert, S., \& van Uitert, E.\ 2011, \mnras, 412, 2095 





\bibitem[Hetterscheidt et 
al.(2005)]{Hetter05} Hetterscheidt, M., 
Erben, T., Schneider, P., Maoli, R., van Waerbeke, L., 
\& Mellier, Y.\ 2005, \aap, 442, 43 

\bibitem[Kaiser et al.(1995)]{KSB} Kaiser, N., Squires, G., 
\& Broadhurst, T.\ 1995, \apj, 449, 460 



\bibitem[Khiabanian 
\& Dell'Antonio(2008)]{Khiabanian08} Khiabanian, H., \& Dell'Antonio, I.~P.\ 2008, \apj, 684, 794

\bibitem[Koester et al.(2007)]{Koester07MaxBCG} Koester, B.~P., et al.\ 
2007, \apj, 660, 239 

\bibitem[Kubo et al.(2009)]{2009ApJ...702..980K} Kubo, J.~M., Khiabanian, 
H., Dell'Antonio, I.~P., Wittman, D., \& Tyson, J.~A.\ 2009, \apj, 702, 980 


\bibitem[Kurtz \& Mink(1998)]{Kurtz98}
Kurtz, M.~J., \& Mink, D.~J.\ 1998, \pasp, 110, 934 

\bibitem[Leauthaud et al.(2007)]{leauthaud07} Leauthaud, A., et 
al.\ 2007, \apjs, 172, 219 



\bibitem[Lubin et al.(1998)]{Lubin98} Lubin, L.~M., Postman, 
M., Oke, J.~B., Ratnatunga, K.~U., Gunn, J.~E., Hoessel, J.~G., 
\& Schneider, D.~P.\ 1998, \aj, 116, 584 

\bibitem[Lubin et al.(2000)]{Lubin00} Lubin, L.~M., Brunner, 
R., Metzger, M.~R., Postman, M., \& Oke, J.~B.\ 2000, \apjl, 531, L5 



\bibitem[Lubin et al.(2004)]{Lubin04} Lubin, L.~M., Mulchaey, 
J.~S., \& Postman, M.\ 2004, \apjl, 601, L9 



\bibitem[Maturi et al.(2007)]{Maturi07} Maturi, M., Schirmer, M., Meneghetti, M., Bartelmann, M., \& Moscardini, L.\ 2007, \aap, 462, 473 

\bibitem[Maturi et 
al.(2010)]{Maturi10} Maturi, M., Angrick, C., Pace, F., \& Bartelmann, M.\ 2010, \aap, 519, A23 

\bibitem[Mink et al.(2007)]{Mink07} Mink, D.~J., Wyatt, W.~F., 
Caldwell, N., Conroy, M.~A., Furesz, G., 
\& Tokarz, S.~P.\ 2007, Astronomical 
Data Analysis Software and Systems XVI, 376, 249 

\bibitem[Miyazaki et al.(2002)]{Miyazaki02} Miyazaki, S., et al.\ 
2002, \apjl, 580, L97 

\bibitem [Miyazaki et al.(2007)]{Miyazaki07} 
Miyazaki, S., Hamana, T., Ellis, R.~S., Kashikawa, N., 
Massey, R.~J., Taylor, J., 
\& Refregier, A.\ 2007, \apj, 669, 714 

\bibitem[Mobasher et al.(2007)]{Mobasher07} Mobasher, B., et al.\ 
2007, \apjs, 172, 117 


\bibitem[Postman et al.(1998)]{Postman98} Postman, M., Lubin, 
L.~M., \& Oke, J.~B.\ 1998, \aj, 116, 560 





\bibitem[Rines \& Diaferio(2006)]{Rines06} Rines, K., \& Diaferio, A.\ 2006, \aj, 132, 1275 

\bibitem[Rines et al.(2003)]{Rines03} Rines, K., Geller, M.~J., 
Kurtz, M.~J., \& Diaferio, A.\ 2003, \aj, 126, 2152 

\bibitem[Roll et al.(1998)]{Roll98} Roll, J.~B., Fabricant, 
D.~G., \& McLeod, B.~A.\ 1998, \procspie, 3355, 324 

\bibitem[Schirmer et 
al.(2007)]{Schirmer07} Schirmer, M., Erben, T., Hetterscheidt, 
M., \& Schneider, P.\ 2007, \aap, 462, 875 

\bibitem[Schneider(2006)]{schneider06} Schneider, P.\ 2006, 
Gravitational Lensing: Strong, Weak and Micro, Saas-Fee Advanced Courses, 
Volume 33.~ISBN 978-3-540-30309-1.~Springer-Verlag Berlin Heidelberg, 2006, 
p.~269, 269 


\bibitem[Shan et al.(2011)]{2011arXiv1108.1981S} Shan, H., Kneib, J.-P., 
Tao, C., et al.\ 2011, arXiv:1108.1981 



\bibitem[Spergel et al.(2007)]{Spergel07} Spergel, D.~N., et al.\ 
2007, \apjs, 170, 377 

\bibitem[Struble 
\& Rood(1999)]{Struble} Struble, M.~F., \& Rood, H.~J.\ 1999, \apjs, 125, 35 


\bibitem[Taniguchi et al.(2007)]{Taniguchi07} Taniguchi, Y., et 
al.\ 2007, \apjs, 172, 9 

\bibitem[White et al.(2002)]{White02} White, M., van Waerbeke, 
L., \& Mackey, J.\ 2002, \apj, 575, 640 


\bibitem[Williams et al.(1996)]{williams96} Williams, R.~E., et 
al.\ 1996, \aj, 112, 1335 





\bibitem[Wittman et al.(2001)]{2001ApJ...557L..89W} Wittman, D., Tyson, 
J.~A., Margoniner, V.~E., Cohen, J.~G., 
\& Dell'Antonio, I.~P.\ 2001, \apjl, 557, L89 


\bibitem[Wittman et al.(2006)]{Wittman06} Wittman, D., 
Dell'Antonio, I.~P., Hughes, J.~P., Margoniner, V.~E., Tyson, J.~A., Cohen, 
J.~G., \& Norman, D.\ 2006, \apj, 643, 128 



\end{thebibliography}
\end{document}